\documentclass[a4paper,11pt]{article}
\usepackage{amsmath}
\usepackage{amssymb}
\usepackage{graphicx}
\usepackage{cite}

\newlength{\dinwidth}
\newlength{\dinmargin}
\renewcommand{\theequation}{\arabic{section}.\arabic{equation}}

\newcounter{figures}
\newcounter{tables}
\makeatletter
\def\fmslash{\@ifnextchar[{\fmsl@sh}{\fmsl@sh[0mu]}}
\def\fmsl@sh[#1]#2{%
  \mathchoice
    {\@fmsl@sh\displaystyle{#1}{#2}}%f
    {\@fmsl@sh\textstyle{#1}{#2}}%
    {\@fmsl@sh\scriptstyle{#1}{#2}}%
    {\@fmsl@sh\scriptscriptstyle{#1}{#2}}}
\def\@fmsl@sh#1#2#3{\m@th\ooalign{$\hfil#1\mkern#2/\hfil$\crcr$#1#3$}}
\makeatother
\newcommand{\mytab}[4][abcd\thetables]{\refstepcounter{tables}%
                       \begin{table}[thb]%
		       \begin{center}%
		       \begin{tabular}{#2}%
		       #3%
		       \end{tabular}%
		       \end{center}%
		       \caption{\label{#1}\small #4}%
		       \end{table}}

\newcommand{\rep}[1]{(\mathbf{#1})}%
\newcommand{\ME}[2]{\left\langle #1|\rep{#2}|\bar{3} \right\rangle}%

\author{Martin Jung}
\title{Determining weak phases from $B\to J/\psi P$ decays}
\date{June 11, 2012}

\begin{document}
\begin{titlepage}
\begin{flushright} DO-TH 12/15 \end{flushright}
\vskip 2cm
\begin{center}
{\Large\bf Determining weak phases from $\mathbf{B\to J/\boldsymbol{\psi} P}$ decays}
\vskip 1cm
\renewcommand{\thefootnote}{\alph{footnote}}
{\bf Martin Jung}\footnote{email: \texttt{martin2.jung@tu-dortmund.de}}
\setcounter{footnote}{0}
\\[0.5cm]
\emph{Institut f\"ur Physik, Technische Universit\"at Dortmund, D-44221 Dortmund, Germany}
\\[0.5cm]
June 11, 2012
\end{center}
\vskip 2cm

\begin{abstract}
The decay $B\to J/\psi K_S$ remains the most important source of information for the $B_d$ mixing phase, determined by the CKM angle $\beta$ in the standard model. When aiming at a precision appropriate for present and coming high luminosity colliders, the corresponding hadronic matrix elements are a major obstacle, as their precise calculation is still not feasible with existing methods. Flavour symmetries offer a possibility to extract them from data, however again with limited precision. In this article, we propose a framework to take subleading contributions in $B_{u,d,s}\to J/\psi P$ decays into account, $P\in \{\pi,K,(\eta_8)\}$, using an $SU(3)$ analysis, together with the leading corrections to the symmetry limit. This allows for a model-independent extraction of the $B_d$ mixing phase adequate for coming high precision data, and additionally yields information on possible New Physics  contributions in these modes. We find the penguin-induced correction to be small, $|\Delta S|\lesssim0.01$, a limit which can be improved with coming data on CP asymmetries and branching ratios. Finally, the sensitivity on the CKM angle $\gamma$ from these modes is critically examined, yielding a less optimistic picture than previously envisaged.
\end{abstract}
\thispagestyle{empty}
\vfill
\end{titlepage}

\section{Introduction}
With the LHC up and running, we have entered a new era in the search for new physics (NP), as well as in the determination of fundamental parameters of the standard model (SM). In flavour physics, a central role is played by the weak phase entering the CKM matrix, and the $B$ factories 
together with the Tevatron experiments have done an amazing job in confirming the CKM mechanism as the main source of low-energy CP violation. The $\mathcal{O}(\%)$-measurement of the CKM angle $\beta$ has been a highlight in that respect. This precision became possible due to the fact that in the  
``golden mode'', $B_d\to J/\psi K_S$, 
explicit calculation of the relevant matrix elements can be avoided, once subleading doubly Cabibbo suppressed terms are assumed to vanish \cite{Bigi:1981qs},
in combination with a final state with a very clear signature. However, given the precision the LHC experiments and planned next-generation $B$ factories are aiming at for this mode and related ones, a critical reconsideration of the assumptions used is mandatory. 

The endeavour of including subleading contributions has been started already some time ago, see e.g. \cite{FleischerPsiK,Ciuchini2005,Ciuchini:2011kd,FFJM,Faller:2008gt,Jung:2009pb,DeBruyn:2010hh,Lenz:2011zz}. As pointed out there,  
in addition to including these terms, the procedure allows for an improved access to NP in $B$ meson mixing, as the SM correction constitutes a ``fake signal'' in that analysis. This is especially interesting, as for several years different extractions of the corresponding angle $\beta$ in the unitarity triangle (UT) have been only marginally consistent (see e.g. \cite{Deschamps:2008de,FJM,Bona:2009cj,Lunghi:2009ke,Charles:2011va}), especially when including $B\to \tau\nu$ in the fit.
 
The main problem lies in the evaluation of hadronic matrix elements, which still does not seem feasible to an acceptable precision for the decays in question. 
To avoid explicit calculation, typically symmetry relations are used, which again have only limited precision. Furthermore, most of the existing proposals to address this question use (mild) additional dynamical assumptions. In this work, we address both issues: we use symmetry under flavour $SU(3)$ to relate $B\to J/\psi P$ decays, where $P$ stands for a pseudoscalar (octet) meson, without neglecting the subleading contributions; in addition, we model-independently include corrections to this limit, without referring to factorization assumptions. Doing so, we demonstrate the importance of $SU(3)$-breaking contributions in this procedure, whose neglect can lead to an overestimation of the subleading effects. Furthermore, the additional assumption used in \cite{Ciuchini2005,Ciuchini:2011kd,FFJM} 
is discussed in a more quantitative manner, to be able to determine a possible influence from this source as well.
Apart from these points, the decays in question also have some sensitivity 
to the CKM angle $\gamma$, see e.g. \cite{FleischerPsiK,DeBruyn:2010hh}. While these modes are  not competitive in that respect to e.g. $B\to DK$, the extraction could serve as an independent crosscheck, and in the event of a largely deviating value even as a hint of NP in the decay amplitudes.

The outline of this article is as follows: we introduce $B\to J/\psi P$ decays in  section~\ref{sec::BtoJPsiP}, pointing out the main features of the decay amplitudes which facilitate the following analysis, and presenting the available data. In section~\ref{sec::su3}, we perform the $SU(3)$ analysis, including linear breaking terms. The resulting expressions are then used in section~\ref{sec::phenomenology} to perform first fits to the available data, to illustrate the impact future measurements are going to have, and to investigate the potential influence of NP. We conclude in section~\ref{sec::conclusions}.

\section{Non-leptonic $\mathbf{B\to J/\boldsymbol{\psi} P}$ decays\label{sec::BtoJPsiP}}
The main focus of this work
 lies on the decay $B_d\to J/\psi K$, from whose indirect CP asymmetry the $B_d$ mixing phase can be extracted. It is related by flavour symmetry to the group of  modes $B\to J/\psi P$, where $B\in \{B^-,\bar{B}^0,\bar{B}_s\}$, and $P$ denotes a pseudoscalar meson from the same multiplet as the kaon, i.e. $P\in \{K,\pi,(\eta_8)\}$. 
Using the unitarity of the CKM matrix, the relevant SM hamiltonian has the structure (see e.g. \cite{BBL} and references therein)
\begin{eqnarray}
\mathcal{H}&=&\sum_{p=d,s}\sum_{q=u,c}V_{qb}^*V_{qp}\left(\sum_{i=1}^2 C_i^q\mathcal{O}_i^q-\sum_{i=3}^{10}C_i\mathcal{O}_i\right)\nonumber\\
&\equiv&\sum_{p=d,s}\mathcal{H}_c^{b\to p}+\mathcal{H}_u^{b\to p}\equiv\mathcal{H}_c+\mathcal{H}_u\,,
\end{eqnarray}
where $\mathcal{O}_i$ are four-quark operators stemming from tree ($i=1,2$), penguin ($i=3\ldots6$) and electroweak penguin ($i=7\ldots10$) diagrams. 
It describes both classes of decays relevant here, corresponding to a change in strangeness $\Delta S=-1$ or $\Delta S=0$. 
The corresponding $b\to s$ transitions are famous for being theoretically clean \cite{Bigi:1981qs}, i.e. to very good approximation dominated by one hadronic amplitude which cancels in the expression for the indirect CP asymmetry. Estimates  yield corrections of the order $\mathcal{O}(10^{-3})$, only \cite{Boos,Li,GronauRosnersuppressedterms}; it is however notoriously difficult to actually calculate the relevant matrix elements, and non-perturbative enhancements cannot be excluded.  
Given this fact, and the impressive experimental precision the flavour factories and the Tevatron have reached already, and the LHC and future Super Flavour Factories (SFF) are about to achieve, 
subleading contributions should be taken into account \cite{FleischerPsiK,Ciuchini2005,Ciuchini:2011kd,FFJM,Faller:2008gt,Jung:2009pb,DeBruyn:2010hh,Lenz:2011zz}. To this aim, the above mentioned symmetry-related processes are important, especially those with $b\to d$ transitions. For them, the suppression of the subleading contributions is much weaker, as the relative CKM factor is lacking the $\lambda^2$ suppression, $|V_{ub}^*V_{ud}|/|V_{cb}^*V_{cd}|\approx R_u=\sqrt{\bar{\rho}^2+\bar{\eta}^2}\approx0.37$. This gives a handle to constrain them by data directly, making the extraction of the $B_d$ mixing phase $\phi$ possible without dynamical assumptions. However, the main amplitudes in these 
 transitions are Cabibbo-suppressed compared to their $b\to s$ counterparts, making the experimental information scarce. Nevertheless, the high statistics measurements becoming available 
make a future precision analysis feasible.

To fully exploit the available information, the approach proposed here is to make a full $SU(3)$ analysis, and thus to include 5-6 different decay modes\footnote{We will consider modes including singlets, i.e. $B_{d,s}\to J/\psi \eta^{(')}$, in a future publication \cite{JS:2012xx}.}. This has the additional advantage to allow for the model-independent inclusion of the leading $SU(3)$-breaking contributions, albeit at the price of a slightly more complicated analysis. These contributions  turn out to be crucial for a reliable extraction of the correction we are aiming at.  
Finally, the extraction of the CKM angle $\gamma$ proposed in \cite{FleischerPsiK} is included and generalized automatically, such that we can investigate its precision in the presence of $SU(3)$-breaking corrections.
While  
different aspects of this anaysis 
have been discussed to more or less extent in the past \cite{Zeppenfeld,FleischerPsiK,Ciuchini2005,Ciuchini:2011kd,FFJM,Jung:2009pb,DeBruyn:2010hh}, 
the combination and extension proposed here 
is critical 
to achieve the desired precision for the decays in question in a model-independent manner.

The experimental data we are going to use are given in Tab.~\ref{tab::expdata}.
\mytab[tab::expdata]{|l|c c c c|}{\hline
Decay                           & $BR/10^{-4}$     & $A_{\rm CP}/\%$ & $S_{\rm CP}$    & Ref.\\\hline
$\bar{B}^0\to J/\psi \bar{K}^0$ & $8.71\pm0.32$    & $0.6\pm 2.1$    & $0.665\pm0.022$ & \cite{Nakamura:2010zzi,Asner:2010qj}\\
$\bar{B}^0\to J/\psi \pi^0$     & $0.176\pm 0.016$ & $10\pm13$       & $-0.93\pm0.15^*$  & \cite{Asner:2010qj,Nakamura:2010zzi}\\
$B^-\to J/\psi K^-$             & $10.13\pm0.34$   & $0.1\pm0.7$     & ---             & \cite{Nakamura:2010zzi}\\
$B^-\to J/\psi \pi^-$ (WA)      & $0.50\pm0.04$    & $1\pm 7$        & ---             & \cite{Nakamura:2010zzi}\\
dataset 2 (LHCb)                & $0.39\pm0.02$    & $0.5\pm2.9$     & ---             & \cite{Aaij:1432547}\\
$\bar{B}^s\to J/\psi K^0$       & $0.33\pm0.04$    &                 &                 & \cite{Aaltonen:2011sy,Aaij:2012xx}\\\hline
}
{Experimental data used for the analysis. The values used from HFAG are from after the Moriond 2012 update. For $B^-\to J/\psi\pi^-$ the first line corresponds to the former world average data (dataset 1), the second to the new LHCb results (dataset 2), as explained in the text. ${}^{*}$ This value is taken from HFAG, without enhancing the error due to  differences of the two averaged measurements.\label{tab::expinput}}
Specifically two recent measurements are noteworthy:
The ratio of branching fractions
\begin{equation}
R_{KK}\equiv\frac{BR(\bar{B}_s\to J/\psi K_S)}{BR(\bar{B}^0\to J/\psi K_S)}
\end{equation}
has been measured by the CDF- and LHCb-collaborations \cite{Aaltonen:2011sy,Aaij:2012xx}; their values agree well with each other, leading to the average given below. This provides the first input from a $B_s$ decay to this analysis.
Furthermore, 
the ratio
\begin{equation}
R_{\pi K}\equiv\frac{BR(B^-\to J/\psi \pi^-)}{BR(B^-\to J/\psi K^-)}
\end{equation}
has been measured by the LHCb collaboration \cite{Aaij:1432547} about 3 standard deviations away from the former world average of all direct measurements of this quantity. 
Instead of averaging the two results we will perform fits  with each of them individually to quantify their influence. 
We call dataset 1 the one including the previous world averages for $R_{\pi K}$ and $A_{\rm CP}(B^-\to J/\psi \pi^-)$, and  dataset 2 the one with the recent measurements of LHCb instead. 

Apart from these recent developments, 
the most striking feature of the data is the com\-pa\-ti\-bility of all direct CP asymmetries with zero, constraining mainly the imaginary parts of the introduced strong amplitudes. For the CP asymmetries, we use the notation
\begin{eqnarray}
a_{\rm CP}(t) &\equiv & \frac{\Gamma(\bar{B}_{d,s}\to J/\psi X^0)-\Gamma(B_{d,s}\to J/\psi X^0)}{\Gamma(\bar{B}_{d,s}\to J/\psi X^0)+\Gamma(B_{d,s}\to J/\psi X^0)}\nonumber\\
&=& \frac{S(B_{d,s}\to J/\psi X^0) \sin(\Delta m_{d,s}t)+A_{\rm CP}(B_{d,s}\to J/\psi X^0) \cos(\Delta m_{d,s}t)}{\cosh(\Delta\Gamma_{d,s}t/2)-A_{\Delta\Gamma_{d,s}}(B_{d,s}\to J/\psi X^0)\sinh(\Delta\Gamma_{d,s}t/2)}\,,
\end{eqnarray}
where $X^0=\{K_{S,L},\pi^0\}$, $S(B_{d,s}\to J/\psi X^0)$ and $A_{\rm CP}(B_{d,s}\to J/\psi X^0)$ are referred to as indirect and direct CP asymmetry, respectively, and $S(B_{d,s}\to J/\psi X^0)^2+A_{\rm CP}(B_{d,s}\to J/\psi X^0)^2+A_{\Delta\Gamma_{d,s}}(B_{d,s}\to J/\psi X^0)^2=1$ holds as well as $\Delta\Gamma_d\approx0$.

Another interesting observation is the relatively large central value of the rate difference
\begin{equation}
A_I^K\equiv\frac{\bar{\Gamma}(\bar{B}^0\to J/\psi \bar{K}^0)-\bar{\Gamma}(B^-\to J/\psi K^-)}{\bar{\Gamma}(\bar{B}^0\to J/\psi \bar{K}^0)+\bar{\Gamma}(B^-\to J/\psi K^-)}\simeq -0.037\pm0.025\,.
\end{equation}
While this measurement is still compatible with zero, a confirmation of this central value would imply much larger contributions from isospin-changing ($\Delta I=1$)  operators than expected in the SM. 

A more extensive discussion of these observables and their relation to the parameters in question is given in Sec.~\ref{sec::phenomenology}.
Note that, although the branching ratios are given explicitly in Tab.~\ref{tab::expdata}, we use in the numerical analysis the measured ratios where available, in order not to correlate the data additionally. We use the values $R_{KK}=0.038\pm0.004$, $R_{\pi K}^{\rm set1}=0.049\pm0.004$, and $R_{\pi K}^{\rm set2}=0.0383\pm0.0013$. For $R_{KK}$ we apply the correction due to the different definitions typically used in theory and experiment, recently discussed in \cite{deBruyn:2012wj}\footnote{In the extraction of the correction, the authors use exact $SU(3)$ symmetry. However, the assigned error is large enough to accommodate also deviations from that limit.} and also applied in \cite{Aaij:2012xx}.

\section{$\mathbf{SU(3)}$ analysis \label{sec::su3}}
In this section, 
the expressions for the decay amplitudes including penguin  and leading first order $SU(3)$-breaking contributions are given. 
We concentrate here on the main results of that analysis; technical details are relegated to appendix~\ref{sec::su3app}.

We start by performing the analysis in the $SU(3)$ limit.  
For $\mathcal{H}_c$, we observe that neglecting the $\Delta I=1,3/2$ part of the electroweak-penguin operators, the flavour structure amounts to a pure triplet,
\begin{equation}
\mathcal{H}_c^{b\to s}\sim \rep{3}_{0,0,-2/3}\,,\quad{\rm and}\quad \mathcal{H}_c^{b\to d}\sim \rep{3}_{1/2,-1/2,1/3}\,,
\end{equation}
using the notation $\rep{R}_{I,I_z,Y}$ for a representation $\mathbf{R}$ with isospin $I$ and hypercharge $Y$.  
The neglected contributions 
are suppressed by small Wilson coefficients, and additionally 
by the fact that these matrix elements do not receive factorizable contributions, which we will call ``dynamical suppression'' in the following. We deem the combined suppression to be strong enough to neglect this contribution, even when 
considering $SU(3)$-breaking contributions.

For $\mathcal{H}_u$, the situation is complicated by the $\Delta I=1,3/2$ part of the occuring tree operators, which have large Wilson coefficients and are only weakly CKM suppressed in $b\to d$ decays.
They are however again dynamically suppressed.
With these terms, the $SU(3)$ analysis yields three independent reduced matrix elements. We confirm the expressions obtained in \cite{Zeppenfeld}. For a translation of the notation used there to ours, see appendix~\ref{sec::su3app}.
The estimates in \cite{Boos,Li,GronauRosnersuppressedterms} correspond to a ratio $r_{\rm pen}\leq 8\%$ of these matrix elements with the leading one. However, as we allow conservatively for their enhancement, we will consider ratios up to $50\%$, and in some cases even larger enhancements for illustration purposes.

In $\bar{B}_s\to J/\psi \pi^0$, the otherwise 
leading part 
of the amplitude  vanishes due to isospin conservation; the only contributions to this decay are from so-called
penguin-annihilation and exchange topologies, which are expected to exhibit additional suppression $\sim \Lambda_{\rm QCD}/m_b$,
and are usually neglected (see e.g. \cite{Ciuchini2005,Ciuchini:2011kd,FFJM}). 
We discuss the possible inclusion of these terms in Sec.~\ref{sec::phenomenology}.

As the next step we include linear $SU(3)$ breaking due to the finite light-quark mass differences. Neglecting isospin breaking, the relevant tensor structure is 
$\rep{8}_{0,0,0}$, driven by the strange-quark mass term.
The tensor product of the leading-order hamiltonian $\mathcal{H}_c$ with this breaking term leads to three more reduced matrix elements.
These include implicitly the expansion parameter expected for this kind of analysis, $\epsilon\sim m_s/\Lambda\sim 20-30\%$, where $\Lambda$ denotes some hadronic scale, as reflected e.g. in the decay constant ratios $(f_K/f_\pi-1)\sim (f_{B_s}/f_{B_d}-1)\sim20\%$. Without further dynamical information, this is also the estimate for their ratios with the leading matrix element, characterized by $r_{SU(3)}$ as defined in the appendix. To be conservative, we allow for $r_{SU(3)}\sim40\%$  in the following fits ($60\%$ for illustration). Note that $r_{SU(3)}\sim1$ corresponds to the absence of any hierarchy, i.e. the reduction of $SU(3)$ symmetry to isospin. There is no sign in $B$ meson decays for such a drastic behaviour, and the fits discussed later indicate in fact a value of the expected order.
For $A_u$, we can neglect these corrections, as already the leading terms are strongly suppressed. 
To our knowledge, these decompositions are derived here for the first time, only a subset has been considered in \cite{Gronauetal1,Jung:2009pb}.

As $A_c(\bar{B}_s\to J/\psi \pi^0)=0$ was enforced by isospin at leading order, all breaking terms vanish as well, as we do not break isospin symmetry. The coefficients for charged and neutral $B\to J/\psi P$ decays ($P\in \{\pi,K\}$)  
remain pairwise forced to be equal by isospin as well (modulo a factor of $\sqrt{2}$ for the $\pi^0$). Related to that is the fact, that the rank of the relevant coefficient matrix is not maximal: all amplitudes can be  expressed by only two independent combinations, without any approximation,
thereby reducing the number of linearly independent amplitudes needed to describe these decays from seven to six.

The framework developed so far yields a description of the six relevant decay amplitudes in terms of 11 hadronic parameters and the CKM phase $\gamma$, 
which reads  
\begin{eqnarray}
A(\bar{B}^0\to J/\psi\bar{K}^0)&=&
\mathcal{N}\left[1+2R_{\epsilon 1}+ 
\bar{\lambda}^2e^{-i\gamma}\left(R_{u1}+\delta\right)\right]\,,\nonumber\\
A(\bar{B}^0\to J/\psi\pi^0)&=&%:&A_{\pi^0}=
-\bar{\lambda}\frac{\mathcal{N}}{\sqrt{2}}
\left[1-R_{\epsilon 1}-R_{\epsilon 2}
-e^{-i\gamma}\left(R_{u1}-\delta\right)\right]\,,\nonumber\\
A(B^-\to J/\psi K^-)&=&
\mathcal{N}\left[1+2R_{\epsilon 1}+\bar{\lambda}^2e^{-i\gamma}\left(R_{u2}-\delta\right)\right]\,,\nonumber\\
A(B^-\to J/\psi\pi^-)&=&%:&A_{\pi^-}=
-\bar{\lambda}\,\mathcal{N}
\left[1-R_{\epsilon 1}-R_{\epsilon 2}
-e^{-i\gamma}\left(R_{u2}-\delta\right)\right]\,,\nonumber\\
A(\bar{B}_s\to J/\psi K^0)&=&
-\bar{\lambda}\,\mathcal{N}
\left[1-R_{\epsilon 1}+R_{\epsilon2} 
-e^{-i\gamma}\left(R_{u1}+\delta\right)\right]\,,\nonumber\\
A(\bar{B}_s\to J/\psi \pi^0) &=& \frac{\mathcal{N}}{\sqrt{2}}\bar{\lambda}^2e^{-i\gamma}(-2\delta)\,.\label{eq::parametrization}
\end{eqnarray}
Here $\mathcal{N}$ denotes the leading amplitude, including the CKM factor $V_{cb}V_{cs}^*$, $R_{\epsilon1,2}$ combinations of $SU(3)$-breaking matrix elements, $R_{u1,2}$ the dominant subleading contributions, and  $\delta$ the remaining one which is expected to be even smaller. The precise definitions of these parameters in terms of reduced matrix elements are given in the appendix. 
Finally we introduced for brevity $\bar{\lambda}=\lambda(1+\lambda^2/2)$, where $\lambda$ denotes the Wolfenstein parameter, and absorbed in these definitions the CKM factor $R_u=\sqrt{\bar{\rho}^2+\bar{\eta}^2}$ and other numerical factors. 
Allowing, as discussed above,
for $SU(3)$ breaking of up to $r_{\rm SU(3)}=40\%$, see also appendix~\ref{sec::su3app},
and an enhancement of the penguin matrix elements such that they may reach $r_{\rm pen}=50\%$ of the leading amplitude,
we expect the order of magnitude of the parameters to be $|R_{\epsilon1,2}|,|R_{u1,2}|\lesssim10-20\%$, 
and $|\delta|\lesssim 1\%$. 
These values correspond to the combination of the generic scales $r_{SU(3)},r_{\rm pen}$ with numerical factors like the CKM factor $R_u$ and Clebsch-Gordan coeffecients, given in the appendix. For $\delta$, we assumed the dynamical suppression to amount to an additional factor of $\sim 20\%$ for the estimate.

When discussing indirect CP violation, the $B$ mixing phases $\phi$ and $\phi_s$ enter. 
The latter is already constrained to be very small, and will be constrained more effectively by the corresponding golden mode $B_s\to J/\psi\phi$, as it appears here only in suppressed decays. Therefore, we set this phase to its SM value, which affects the predictions for the indirect $CP$ asymmetry in $B_s\to J/\psi K_S$ only marginally. While ultimately these 13 (14) parameters should face 16 measurements of decay rates and CP asymmetries, of these only 11 are available so far. The LHCb experiment is expected to determine the CP asymmetries in $\bar{B}_s\to J/\psi K_S$ soon, albeit with limited precision at first, yielding two more constraints.
Therefore, for the time being, additional assumptions are necessary. It is however possible to make them 
in a transparent and testable way, which is one advantage of the present analysis. 

One obvious choice is the neglect of the $\Lambda/m_b$-suppressed contributions discussed above. A qualitative confirmation of this assumption is given by decays dominated by the same topologies (but \emph{not} related by symmetry), as e.g. $\bar{B}^0\to D^0\phi$, to which an upper limit of $\sim 10^{-5}$ exists \cite{Nakamura:2010zzi}, despite an enhancement factor of $|V_{cb}/V_{ub}|^2$ compared to that contribution in $\bar{B}^0\to J/\psi \pi^0$. 
Apart from this,
here we also have the opportunity to test the contributions within the framework itself, which will be discussed in the next section. Making this assumption yields for the present situation 11 parameters for 11 measurements.

Additionally, given that our main interest is the extraction of the mixing phase, we can use input for the CKM phase $\gamma$ to remove another free parameter.
Unless otherwise stated, we will use $\gamma=(67.1\pm 4.3)^\circ$, extracted from a global fit \cite{CKMfitter}\footnote{Although $S(B\to J/\psi K_S)$ enters this analysis, it does not affect the extraction of $\gamma$ significantly. We expect this input soon to be replaced by  a competitive measurement in tree-level processes.}. We will examine the possibility to determine this angle within the fit below. 
These options
show that the method proposed here is already applicable now and will be even more fruitful in the future. 

Considering non-SM contributions to this class of decays, we have two possible contributions, which in general will both be present: new contributions to the mixing and to the decay amplitude. The former may be parametrized as $\phi\to \phi_{\rm SM}+\Delta\phi_{\rm NP}$, and result in an extracted value for the mixing phase $\phi$ different from its SM value. This way to extract $\phi$ takes into account the penguin ``pollution'', thereby excluding a source for a fake signal, and could then be used in an analysis including further observables from $B$ mixing, as e.g. \cite{Lenz:2006hd,Lenz:2012az}.  A model-independent parametrization of %the latter 
NP in the decay amplitudes
yields too many parameters to be fitted in full generality. While for a specific model flavour structure and relevant operators are usually given, allowing for the corresponding fit, in a model-independent approach we are forced to consider scenarios in which certain contributions are dominant, transforming in a definite way under $SU(3)$. 
For minimal flavour violating scenarios \cite{D'Ambrosio:2002ex,Chivukula:1987py,Hall:1990ac,Buras:2000dm} however, the parametrization given above remains valid, only the interpretation of the extracted parameter values changes, because the relative influence of Wilson coefficients and corresponding matrix elements is different.

\section{Phenomenology\label{sec::phenomenology}}
We now proceed with the analysis of the data shown in Tab.~\ref{tab::expdata}. 
First we introduce combinations of observables which obtain their main contributions from fewer 
parameters, similarly to \cite{FleischerMannel1},  
to understand better their physical interpretation. Then we start fitting the data, beginning with  
the $SU(3)$ limit, to examine the necessity of the breaking terms introduced above. As there are less parameters involved here, we can allow for deviations from the limit 
$\delta\equiv0$, 
to check the compatibility of the data with our assumption. This is followed by an analysis including $SU(3)$ breaking, where this contribution is again set to zero; here we first test the limit of vanishing penguin amplitudes, before allowing for both classes of contributions. 
We continue with possible future scenarios, to illustrate the impact of coming measurements, and finally discuss NP contributions to these decays to some extent.

The software used for the fits is the ``NLopt'' software package \cite{NLopt}, together with the augmented lagrangian \cite{AUGLAG1,AUGLAG2} and  Sbplex/Subplex algorithms \cite{NLopt,Subplex}. The best fit values have been checked additionally with Mathematica.

In order to see what to expect from the fit, we introduce a common power-counting parameter $\xi\sim0.2$ for the appearing quantities, where the assignment corresponds to rounding the logarithms $\log_{\xi}(R_i,\delta)$, in accordance with the estimates given earlier in Sec.~\ref{sec::su3}:
\begin{equation}
\bar{\lambda},|R_{u1,2}|,|R_{\epsilon1,2}|\sim\xi\,,%\quad |R_{u2}|\sim\xi^2\,,
\quad {\rm and}\quad |\delta|\sim\xi^3\,.
\end{equation}
The combinations of observables facilitating the interpretation of the introduced paraemters  
are collected in Tab.~\ref{tab::observables} together with the leading order expressions using the power counting defined above. Estimates of parameters via these relations obviously only hold if the power counting is respected by the fit.

The two isospin-related decay pairs offer the possibility to access the parameters $R_{u1,2}$: the direct CP asymmetries
are to leading order given by the corresponding imaginary parts, and the  rate asymmetries $A_I^K$, defined above, as well as 
\begin{eqnarray}
A_I^\pi&\equiv&\frac{2\bar{\Gamma}(\bar{B}^0\to J/\psi \pi^0)-\bar{\Gamma}(B^-\to J/\psi \pi^-)}{2\bar{\Gamma}(\bar{B}^0\to J/\psi \pi^0)+\bar{\Gamma}(B^-\to J/\psi \pi^-)}
\end{eqnarray}
allow to constrain the real part of the combination $\Delta R_u = R_{u1}-R_{u2}$. As can be seen from Tab.~\ref{tab::obscombinations}, the former constrain the imaginary parts to be (at most) of the expected order, while the rate asymmetries yield a less consistent picure: $A_I^K$ prefers ${\rm Re}(\Delta R_u)$ to be an order of magnitude larger than expected and negative (however with a significance of less than two standard deviations), while $A_I^\pi$ in dataset 2 implies a very small value of at most the expected order, and dataset 1 yields a somewhat larger value, however with a positive sign. Combining these observations shows a slight inconsistency between the data and our theoretical expressions, specifically when considering the sum $A_I^K+\bar{\lambda}^2A_I^\pi$, which is expected to be
safely below one per mill. Experimentally, its central value is several percent, dominated by $A_I^K$, but enlarged in dataset 1 by the fact that $A_I^\pi$ exhibits the ``wrong'' sign. 
If this 
is confirmed to higher precision in the future, it might become an actual hint of NP, making the corresponding measurements very interesting.

\mytab[tab::obscombinations]{|c c c|}{\hline
Observable                                   & LO expression                                                 & Experiment\\\hline
$A_I^K$                                      & $\bar{\lambda}^2\cos\gamma\,[{\rm Re}(R_{u1})-{\rm Re}(R_{u2})]$                & $-0.037\pm0.025$\\
$A_I^\pi$                                    & $-\cos\gamma\,[{\rm Re}(R_{u1})-{\rm Re}(R_{u2})]$                              & $-0.13\pm0.06$\\
                                             &                                                              & $-0.01\pm0.05$\\
$A_{\rm CP}(\bar{B}^0\to J/\psi \bar{K}^0)$  & $2\bar{\lambda}^2\sin\gamma \,{\rm Im}(R_{u1})$              & $\phantom{-}0.010\pm0.012$\\
$A_{\rm CP}(B^-\to J/\psi K^-)$              & $2\bar{\lambda}^2\sin\gamma \,{\rm Im}(R_{u2})$              & $\phantom{-}0.001\pm 0.007$\\
$A_{\rm CP}(\bar{B}^0\to J/\psi \pi^0)$  	 & $-2\sin\gamma \,{\rm Im}(R_{u1})$                			& $\phantom{-}0.10\pm0.13$\\
$A_{\rm CP}(B^-\to J/\psi \pi^-)$            & $-2\sin\gamma \,{\rm Im}(R_{u2})$             				& $\phantom{-}0.01\pm 0.07$\\
                                             &                                                              & $\phantom{-}0.005\pm 0.029$\\
$\Delta A_{\rm CP}$ 						 & $4\sin\gamma\,{\rm{Im}}(R_{u1}R_{\epsilon2})$ 				& \phantom{-}---\\
$\Delta S_{J/\psi K}$                    	 & $-2\bar{\lambda}^2\sin\gamma\cos(\phi)\,{\rm Re}(R_{u1})$    & \phantom{-}---\\
$S(B\to J/\psi \pi)+S(B\to J/\psi K)$        & $2\sin\gamma\cos(\phi)\,{\rm Re}(R_{u1})$                    & $-0.27\pm0.15$\\
$\tilde{R}_{\pi K}-\bar{\lambda}^2$   		 & $-2\bar{\lambda}^2\,\left[\cos\gamma\,{\rm Re}(R_{u2})+{\rm Re}(3R_{\epsilon1}+R_{\epsilon2})\right]$                                 & $-0.006\pm0.004$\\
                                             &                                                               & $-0.016\pm0.001$\\
$\tilde{R}_{K K}-\bar{\lambda}^2$ & $-2	\bar{\lambda}^2\,\left[\cos\gamma\,{\rm Re}(R_{u1})+{\rm Re}(3R_{\epsilon1}-R_{\epsilon2})\right]$ & $-0.015\pm0.004$
\\\hline
}
{\label{tab::observables}Combinations of observables to disentangle the contributions of the different parameters, together with their leading order (LO) expressions using the power counting discussed in the text. When two experimental values are present, the one for dataset 2 is given in the second line.}

The $SU(3)$-breaking parameters $R_{\epsilon1,2}$ are best determined by branching ratio combinations. The decay rate difference  
\begin{equation}
A^0_{\pi K}\equiv\frac{\Gamma(\bar{B_s}\to J/\psi K^0)-2\Gamma(\bar{B}^0\to J/\psi \pi^0)}{\Gamma(\bar{B_s}\to J/\psi K^0)+2\Gamma(\bar{B}^0\to J/\psi \pi^0)}\simeq 2{\rm Re}(R_{\epsilon2})
\end{equation}
basically determines ${\rm Re}(R_{\epsilon2})$ to be at most of the expected order, while 
 $\tilde{R}_{\pi K}=\Gamma(B^-\to J/\psi\pi^-)/\Gamma(B^-\to J/\psi K^-)$ and $\tilde{R}_{KK}=\Gamma(\bar{B}_s\to J/\psi K^0)/\Gamma(\bar{B}^0\to J/\psi \bar{K}^0)$ then mainly constrain ${\rm Re}(R_{\epsilon1})$. 
An important point is that the value of $\tilde{R}_{KK}$ (and the one of $\tilde{R}_{\pi K}$ in dataset 2) 
can easily be explained by 
$R_{\epsilon1}$, while the assumption of exact $SU(3)$ symmetry would imply a very large real part of $R_{u1}$ ($R_{u2}$ for $\tilde{R}_{\pi K}$), as their relative coefficient is $3/\cos\gamma\sim8$. 
${\rm Re}(R_{u1})$ 
determines at leading order in our expansion the differences
\begin{equation}
\Delta S_X\equiv\eta_X\,S(B^0\to X)+\sin\phi\,,
\end{equation}
we are aiming to determine.
Here it becomes therefore obvious that assumptions regarding $SU(3)$ breaking influence the extraction of these differences strongly, which is one of the motivations for the model-independent treatment of these terms in the present analysis. The mixing phase can be estimated from the relation
\begin{equation}
-(\eta_{J/\psi K}S(B_d\to J/\psi K_S)+\bar{\lambda}^2\eta_{J/\psi\pi}S(B_d\to J/\psi \pi^0)\simeq(1+\bar{\lambda}^2)\sin\phi\stackrel{\rm exp}{=}0.715\pm0.023\,,
\end{equation}
where $\eta_X$ denotes the CP eigenvalue of the final state $X$, and the main corrections cancel, as can be seen again from Tab.~\ref{tab::observables}.

A problem from the fitting point of view lies in the circumstance that none of the measured observables shown in this list depends on the imaginary parts of $R_{\epsilon1,2}$ on leading order. For that reason these imaginary parts 
are not well determined by the fit. For the same reason, however, predictions of observables will not be strongly influenced by them. In order for these parameters not to make the fit unstable, we constrain them to lie in a reasonable range, which we conservatively choose to correspond to $r_{SU(3)}=40\%$ of the leading matrix element, i.e. $|{\rm Im}(R_{\epsilon1})|\leq 0.1$ and $|{\rm Im}(R_{\epsilon2})|\leq 0.14$, which should affect only unphysical solutions. Another implication of this observation is, that these parameters do not effectively serve to fit the data, i.e. they should not be taken into account when determining the number of degrees of freedom in the fit. Here we simply reduce the number of parameters by two, which is again an approximation, but a reasonable one, given the situation that even huge values of these parameters are allowed. To remind the reader of this, we speak below of \emph{effective degrees of freedom}. 
This situation could improve if the
combination 
\begin{equation}
\Delta A_{\rm CP}=A_{\rm CP}(\bar{B}_s\to J/\psi K^0)-A_{\rm CP}(\bar{B}^0\to J/\psi \pi^0)
\end{equation} 
was measured to high precision, see once more Tab.~\ref{tab::observables}.

\subsection{Fit in the $\mathbf{SU(3)}$ limit}
\begin{figure}
\begin{center}
\parbox{7.5cm}{\includegraphics[width=7.5cm]{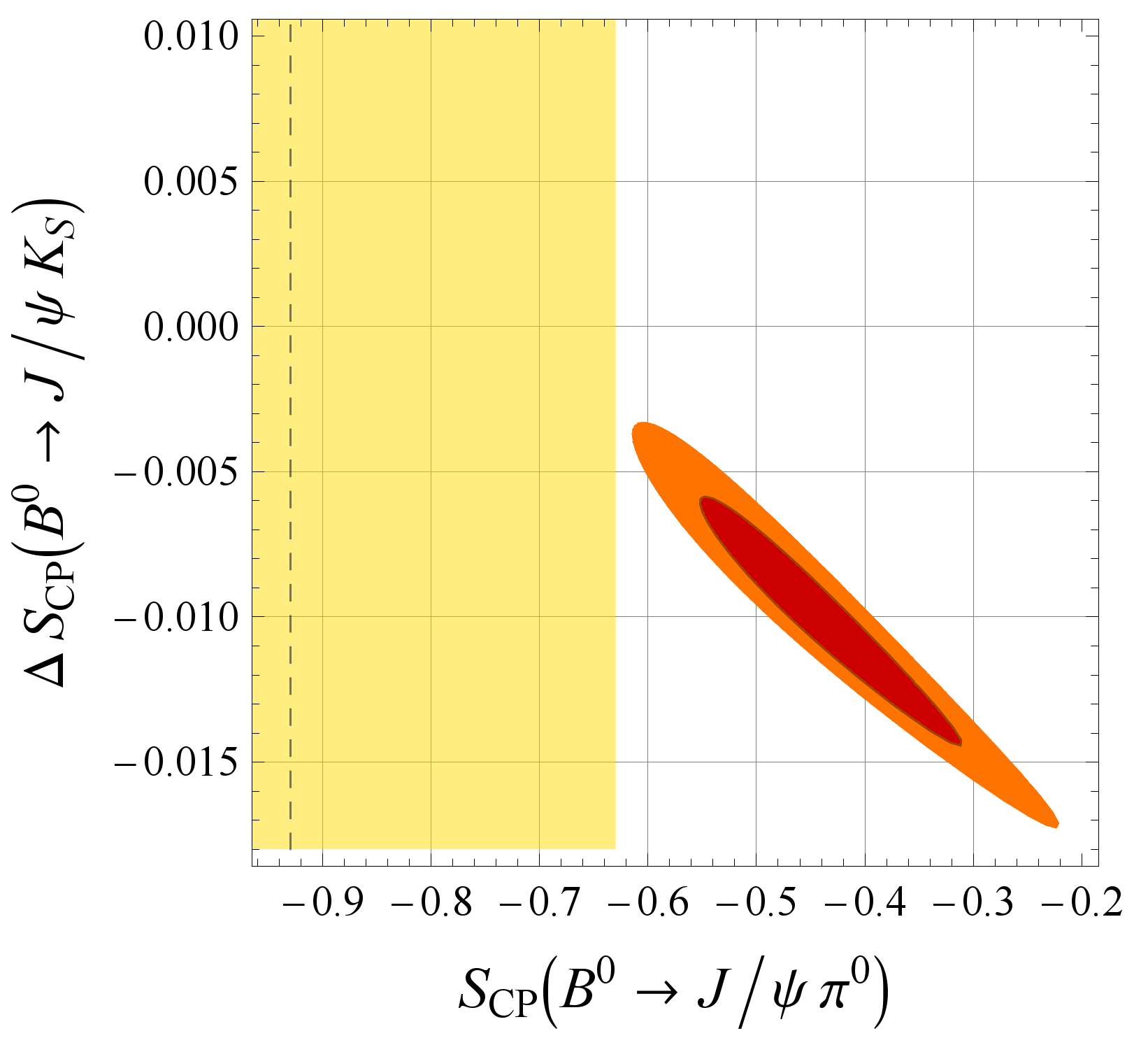}}\phantom{x}\hfill\parbox[b]{7cm}{
\caption{\label{fig::su3fit} Fit result for $\Delta S_{J/\psi K_S}$ versus $S(B_d\to J/\psi\pi^0)$ in the $SU(3)$ limit for dataset~2, ($\chi^2_{\rm min}=23.9$), allowing for arbitrary values of the penguin matrix elements. The inner (outer) area correspond to $68(95)\%$~CL. Note the experimental $2\sigma$ range for $S(B\to J/\psi \pi^0)$, indicated by the light yellow (light grey) box.}}
\end{center}
\end{figure}

Performing the fit in the $SU(3)$ limit ($R_{\epsilon1,2}\to0$), and with $\delta=0$  
yields a rather bad fit with $\chi^2_{\rm min}/{\rm d.o.f.}=22.3(23.9)/5$ for dataset 1(2), even when allowing up to $|R_{u1,2}|=1$. The reasons can be understood from the last paragraph, as the different estimates contradict each other in this limit. The situation worsens once reasonable values for the subleading matrix elements are enforced. Allowing for up to $r_{\rm pen}=50\%$
yields $\chi^2_{\rm min}/{\rm d.o.f.}=25.9(58.5)/5$, where the latter result shows the strong influence of $R_{\pi K}$. The outcome of such a fit is shown exemplarily in Fig.~\ref{fig::su3fit}, where the fit result for $\Delta S_{J/\psi K_S}$ versus $S(B\to J/\psi \pi^0)$ is plotted; we emphasize that this is just for illustration. 
However, this plot allows for two connected interesting observations: the predicted sign of the penguin-induced phase shift is opposite to that in \cite{FFJM}, and $S_{\rm CP}(B\to J/\psi \pi^0)$ is predicted to be much smaller than the present central value. Both results are due to the precisely  measured branching ratios entering the fit and will change again once $SU(3)$ breaking is taken into account. 

We also repeat the fit, taking factorizable $SU(3)$ breaking into account via the corresponding phase space and form factors, as this is advocated sometimes in the literature. This does not improve the situation, but yields $\chi^2_{\rm min}=28.0(23.9)$ for the two datasets. These decays are known anyway to be badly described by factorization, which is why we will not apply any such factors in the following.

We proceed to investigate the influence of the parameter $\delta$ neglected so far. Its inclusion does not improve the situation, either: the new fit yields a $\chi^2_{\rm min}/{\rm d.o.f.}=18.2(20.4)/3$, which is worse than with $\delta=0$. 
We take this observation as confirmation that the limit $\delta\to0$ is reasonable,   
and use this approximation in the following.

\subsection{Fits including $\mathbf{SU(3)}$ breaking}
\begin{figure}
\begin{center}
\includegraphics[width=7.6cm]{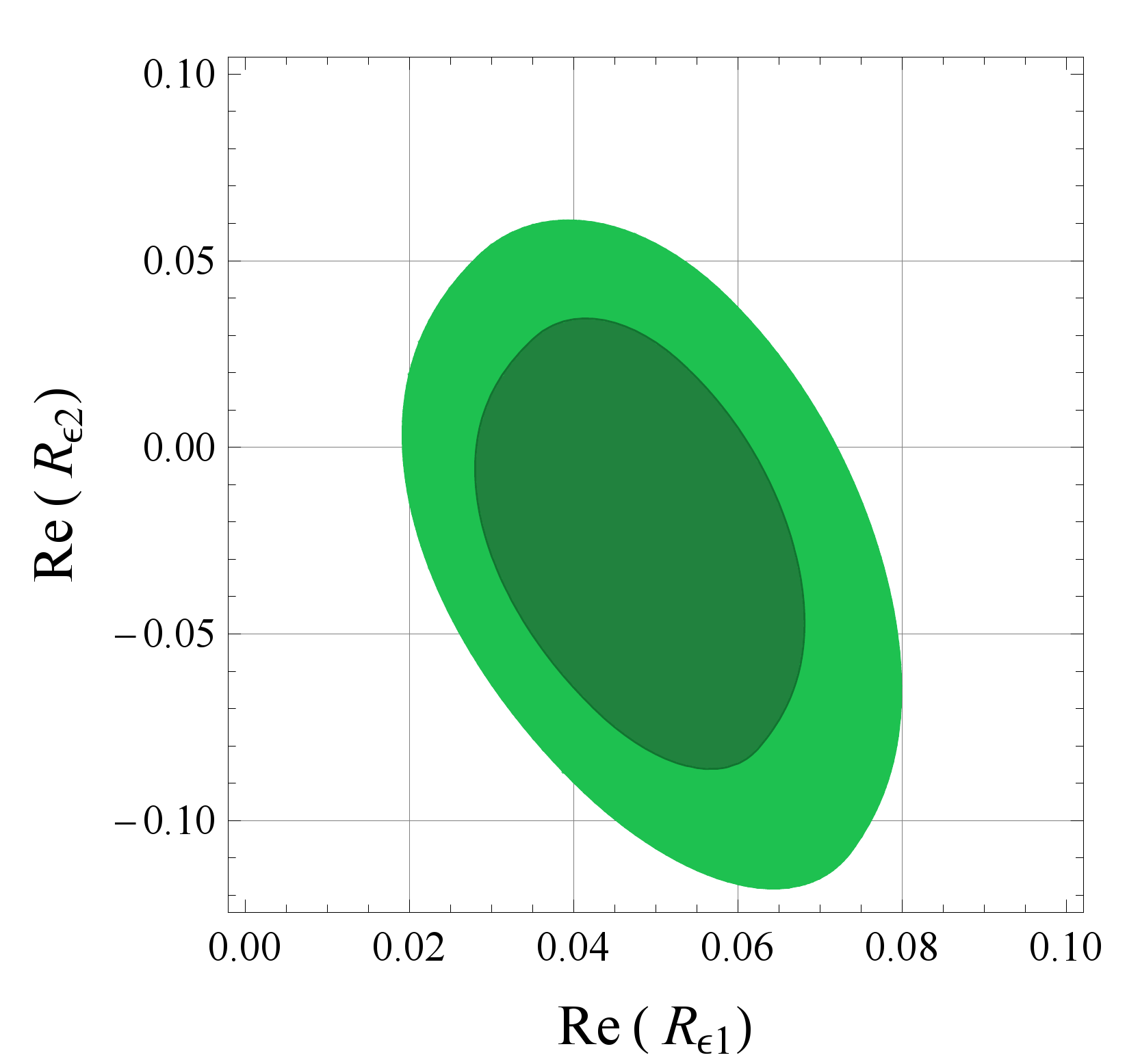}\hfill \includegraphics[width=7.6cm]{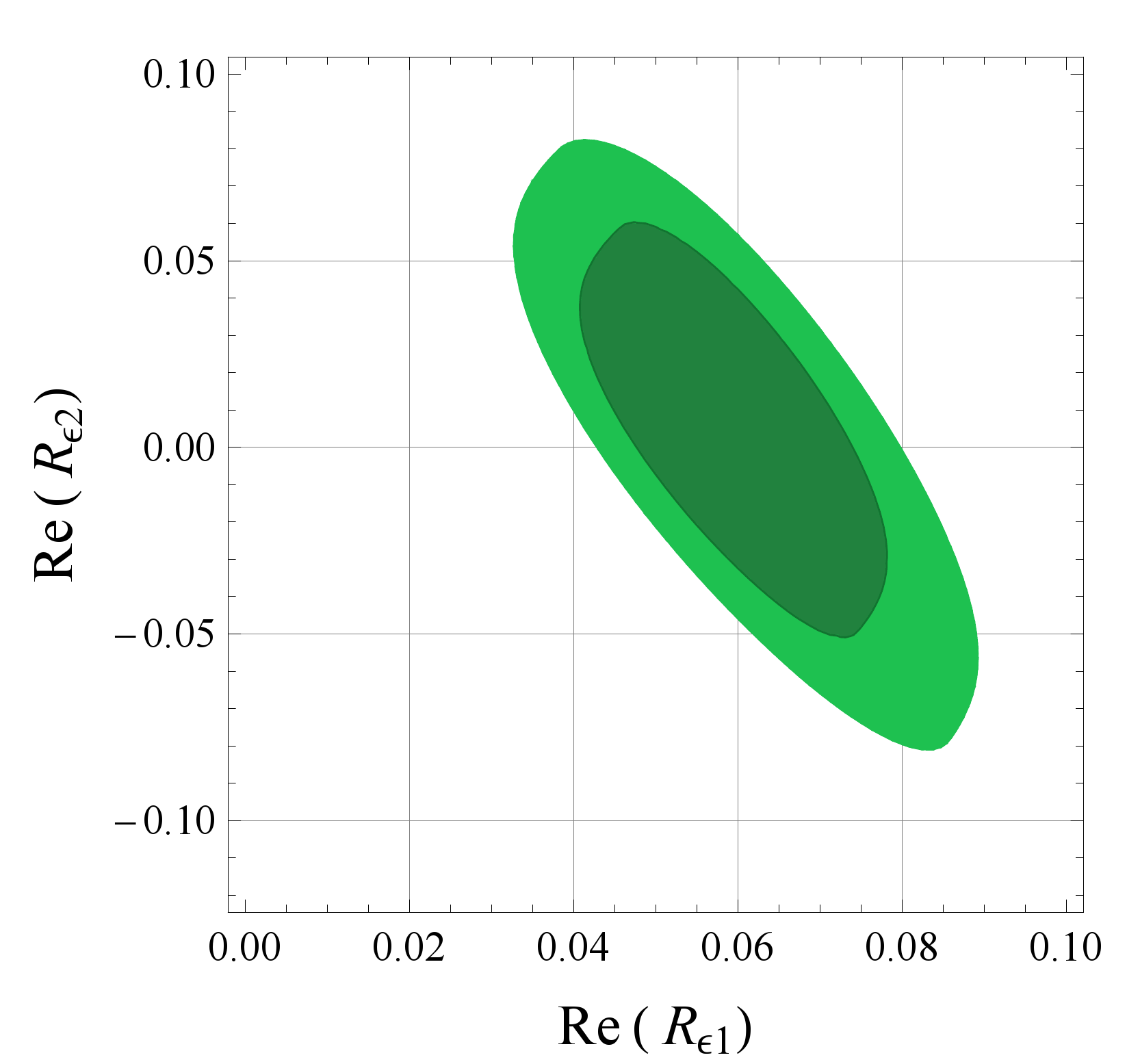}
\caption{\label{fig::resulteo} Fit results ($68/95\%$~CL) in the ${\rm Re}(R_{\epsilon1})-{\rm Re}(R_{\epsilon2})$-plane for datasets 1 (left) and 2 (right), in the limit of vanishing penguin contributions.}
\end{center}
\end{figure}

Including $SU(3)$ breaking, the situation changes significantly. 
In a first step, we  
assume the subleading contributions $\sim V_{ub}^*V_{cb}$ to vanish identically. Obviously this implies vanishing direct CP asymmetries and the indirect ones to be equal to $\pm\sin\phi$, such that the two measurements are effectively averaged to $\sin\phi=0.671\pm0.022$. This fit works quite (very) well for the two datasets, we obtain $\chi^2_{\rm min}=9.4(6.0)$ for 7 effective degrees of freedom. Not to count the imaginary parts of $R_{\epsilon i}$ as parameters is supported by the fact that setting them to zero does not change these values. We show the results in Fig.~{\ref{fig::resulteo}} therefore in the ${\rm Re}(R_{\epsilon1})-{\rm Re}(R_{\epsilon2})$-plane, allowing the imaginary parts to reach values corresponding to $40\%$ of the leading order matrix element. As can be seen there, both fits allow additionally for ${\rm Re}(R_{\epsilon2})=0$, while the best fit corresponds to ${\rm Re}(R_{\epsilon1})=0.047(0.059)$ respectively, yielding a value for the corresponding combination of $SU(3)$-breaking matrix elements  of $19(24)\%$ of the leading order one, which is perfectly within expectations. 
Therefore the data, strictly speaking, do not call for the inclusion of the penguin contributions at the moment (at least for dataset 2), although of course they are not perfectly described without them. In any case more precise measurements of the relevant CP asymmetries will decide the size  of these terms in the future.

Including again the parameters $R_{u1,2}$, we obtain $\chi^2_{\rm min}=2.8(2.3)$ for 3 effective degrees of freedom for the two datasets, when we refrain from applying strong restrictions on the parameter values\footnote{We do not allow for ``exchanging roles'' though, i.e. we continue to assume that $N_0$ represents the leading matrix element.}.   
For this fit, the parameters $R_{\epsilon1,2}$ allow to accommodate the pattern of branching ratios, while the penguin contributions are mainly determined by the CP and isospin asymmetries. The central values of the absolute values $|R_{u1,2}|$ still tend to larger values than theoretically expected. This is not surprising, given the discussion on $A_I^{K,\pi}$ above. The fit shows  
explicitly that the remaining  
$\chi^2_{\rm min}$ is mainly due to the branching ratios in $B\to J/\psi K$: the fitted central values are one standard deviation higher (lower) for $\bar{B}^0\to J/\psi \bar{K}^0\,(B^-\to J/\psi K^-)$,  
underlining the importance of a new measurement of the ratio of these branching ratios, which correspondingly is predicted to take a significantly different central value than the one presently measured. 

Restricting the fit parameters  to the expected ranges shows a difference between the two datasets: it does not worsen the fit as much as in the $SU(3)$ case, but while for dataset 2 it stays very close to the absolute minimum, $\chi_{\rm min, constr.}^{2,{\rm set\,2}}=2.8$, it doubles for dataset 1 to $\chi_{\rm min, constr.}^{2,{\rm set\,1}}=5.6$. The new result for $R_{\pi K}$ obtained by LHCb seems therefore favoured by this fit. While  it is too early to draw  
conclusions, this observation demonstrates once more the importance of precise branching ratio measurements in this context. 
The remaining difference stems from 
slight tensions  
in the $B\to J/\psi\pi^0$ CP asymmetries for dataset 1, which are both predicted to lie $\sim 1\sigma$ below their present central values.  

For both datasets, the shift 
$\Delta S_{J/\psi K}$
now tends again to positive values, thereby lowering the corresponding tension in the UT fit \cite{Deschamps:2008de,FJM,Bona:2009cj,Lunghi:2009ke,Charles:2011va}; it is however still compatible with zero, in agreement with the observation above of a reasonable fit without penguin terms. The obtained ranges read
\begin{eqnarray}
\Delta S_{J/\psi K}^{\rm set\,1} &=& [0.001,0.005] ([-0.004,0.011])\,,\mbox{\quad and }\\
\Delta S_{J/\psi K}^{\rm set\,2} &=& [0.004,0.011] ([-0.003,0.012])\,,
\end{eqnarray}
for $68\%$ ($95\%$) CL, respectively.
The reason for the sign change compared to the $SU(3)$ limit lies in the indirect CP asymmetry in $B\to J/\psi\pi^0$, whose absolute value was forced to be smaller than the one in the golden mode before 
(see Fig.~\ref{fig::su3fit}), but is now allowed to drive the shift to positive values. Differently put, neglecting $SU(3)$ breaking drives $\Delta S_{J/\psi K}$ 
to relatively large values with a different sign. $S(B_d\to J/\psi\pi^0)$ is however still predicted by the fit to lie below the present central value of the measurement, thereby supporting the Belle result \cite{Lee:2007wd} over the BaBar one, which indicates a very large value for this observable \cite{Aubert:2008bs}. 
\begin{figure}
\begin{center}
\includegraphics[width=7.8cm]{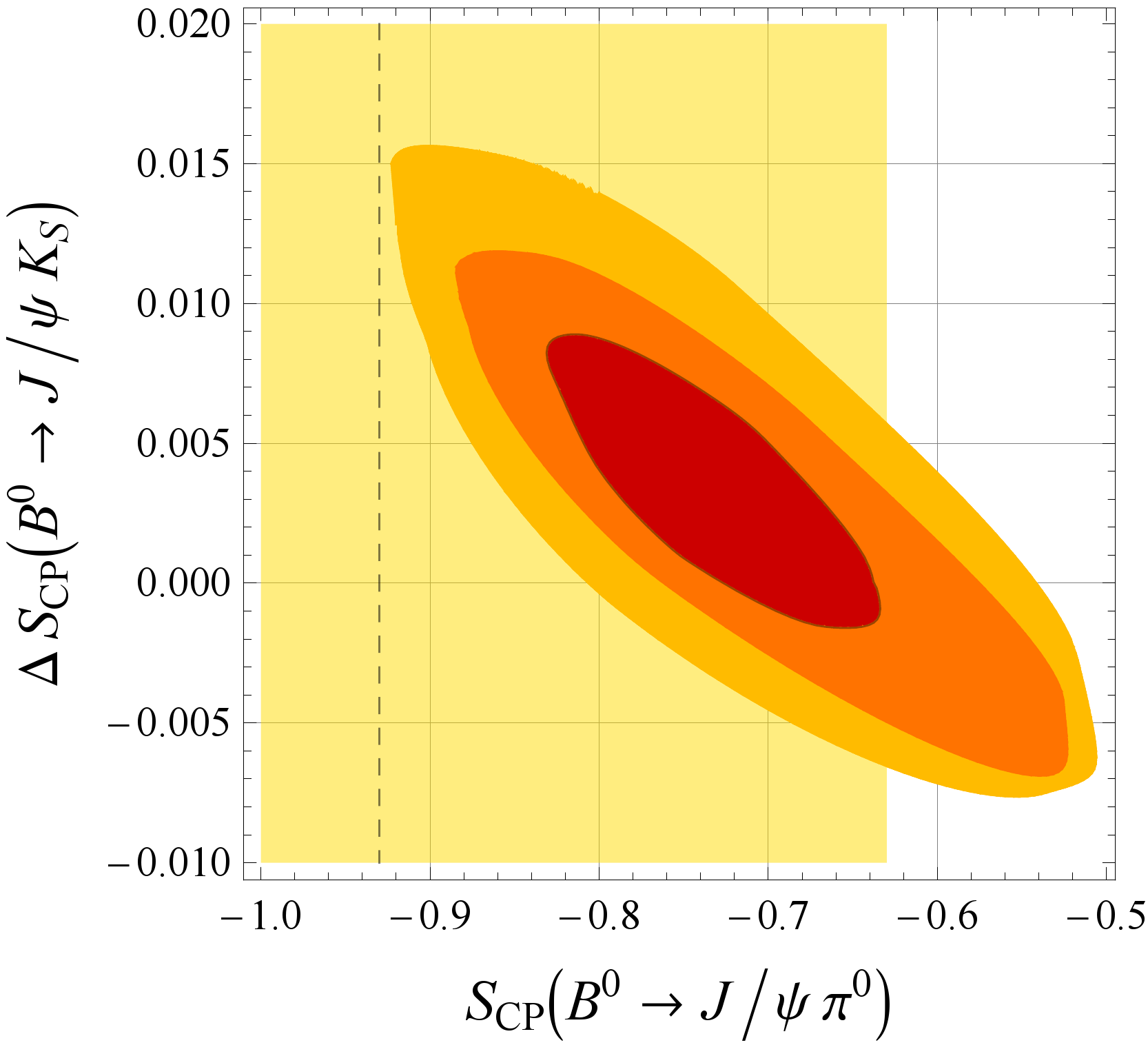}\hfill \includegraphics[width=7.8cm]{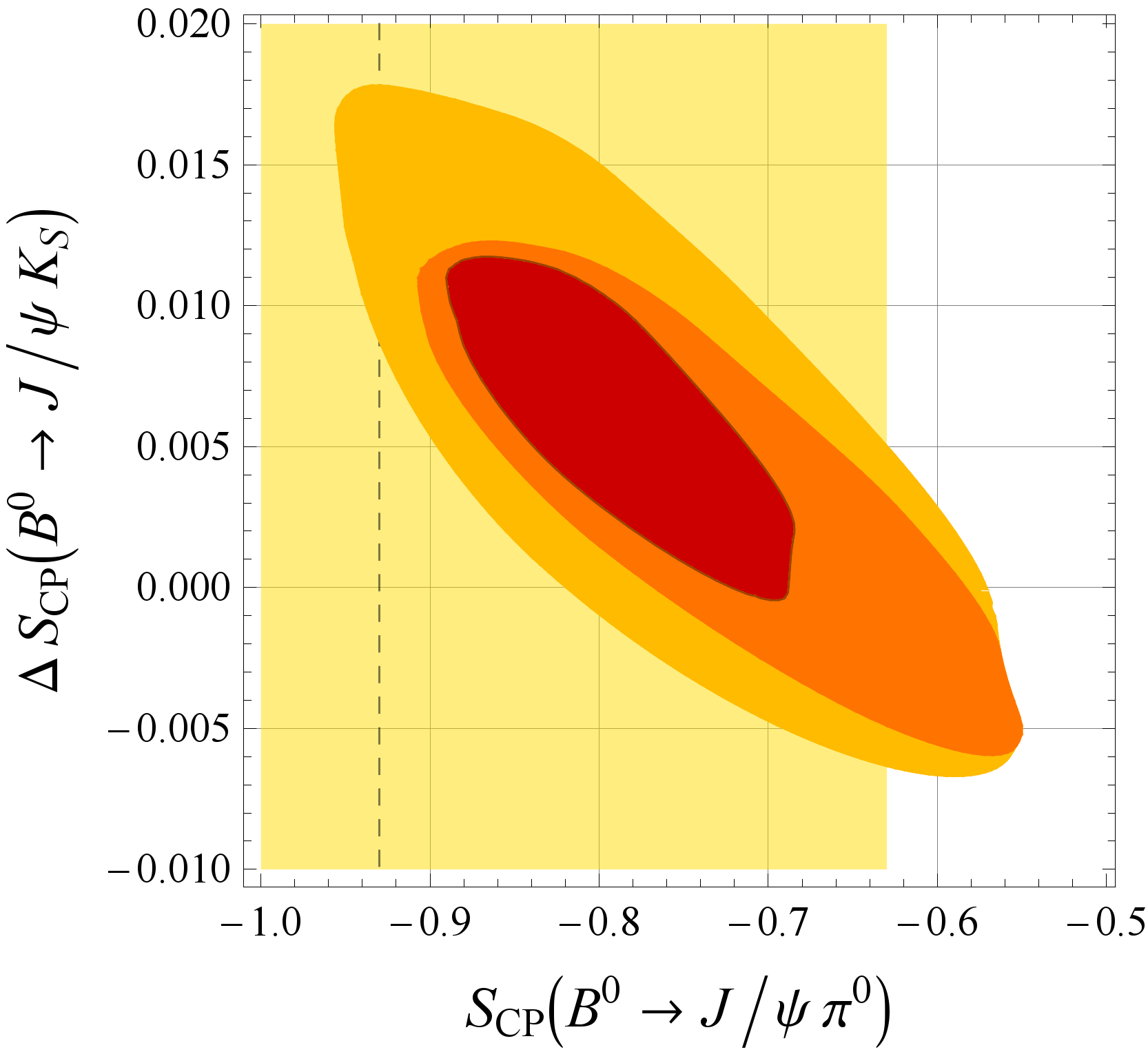}
\caption{\label{fig::resultsfullfita} Fit results for datasets 1 (left) and 2 (right), for $\Delta S_{J/\psi K}$ versus $S_{\rm CP}(B^0\to J/\psi \pi^0)$, including all available data. The inner areas correspond to $68\%$ CL and $95\%$ CL with $r_{SU(3)}=40\%$ and $r_{\rm pen}=50\%$. The outer one is shown for illustration purposes, only, and corresponds to $95\%$ CL when allowing for up to $r_{SU(3)}=60\%$ and $r_{\rm pen}=75\%$. The light yellow area indicates the 2-$\sigma$ range of the $S(B^0\to J/\psi \pi^0)$ average, the dashed line its central value.}
\end{center}
\end{figure}
These findings are illustrated in Fig.~\ref{fig::resultsfullfita}.
The mixing phase is extracted as $\phi_{\rm fit}=0.74\pm0.03$ (for both datasets), which is to be compared with $\phi_{\rm SM}^{\rm naive}=0.73\pm0.03$ when using the naive relation without penguin contributions. The inclusion of the correction therefore yields the same precision, but a slightly different central value.

\begin{figure}[htb]
\begin{center}
\includegraphics[width=7.3cm]{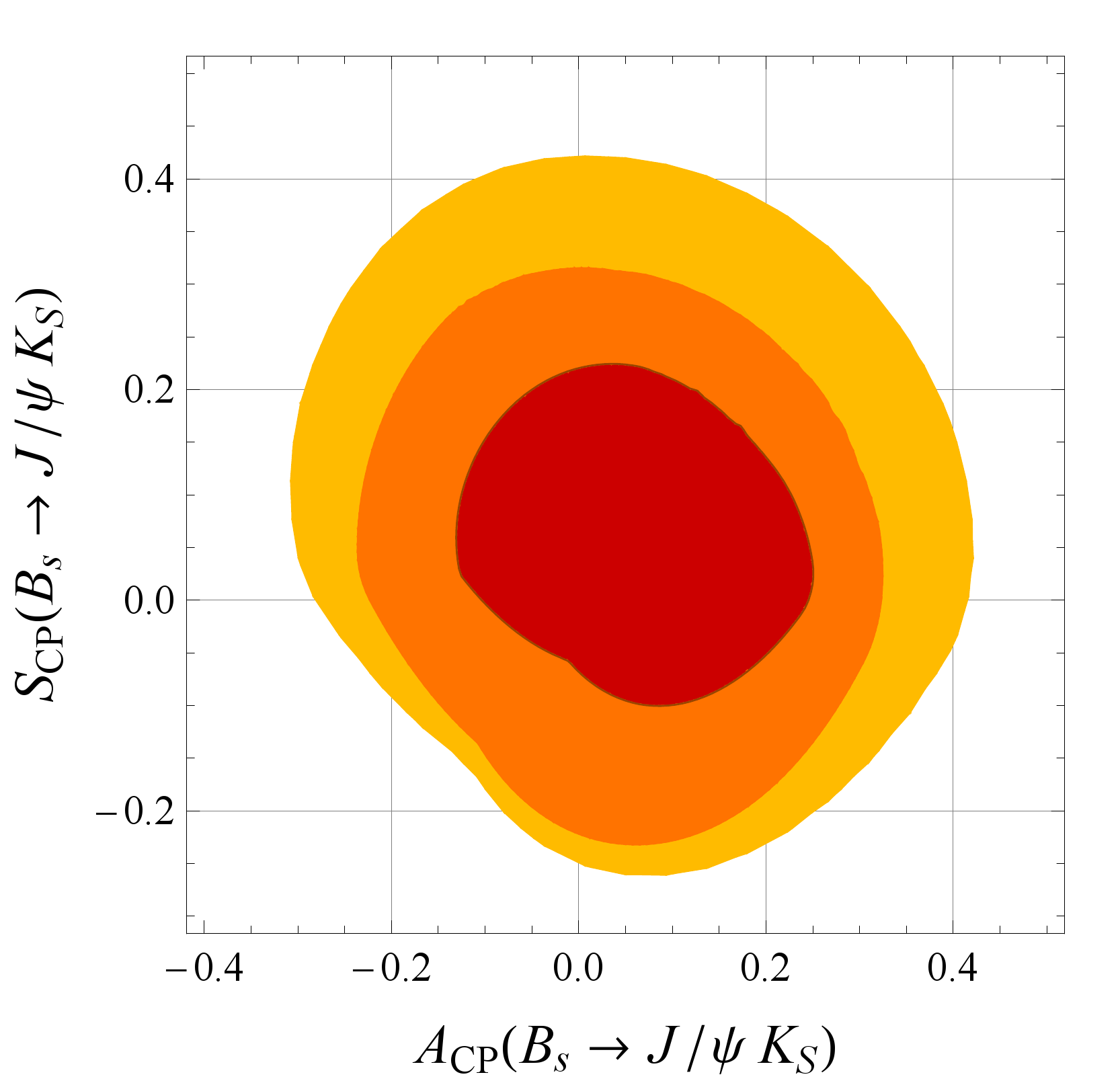}\hfill \includegraphics[width=7.3cm]{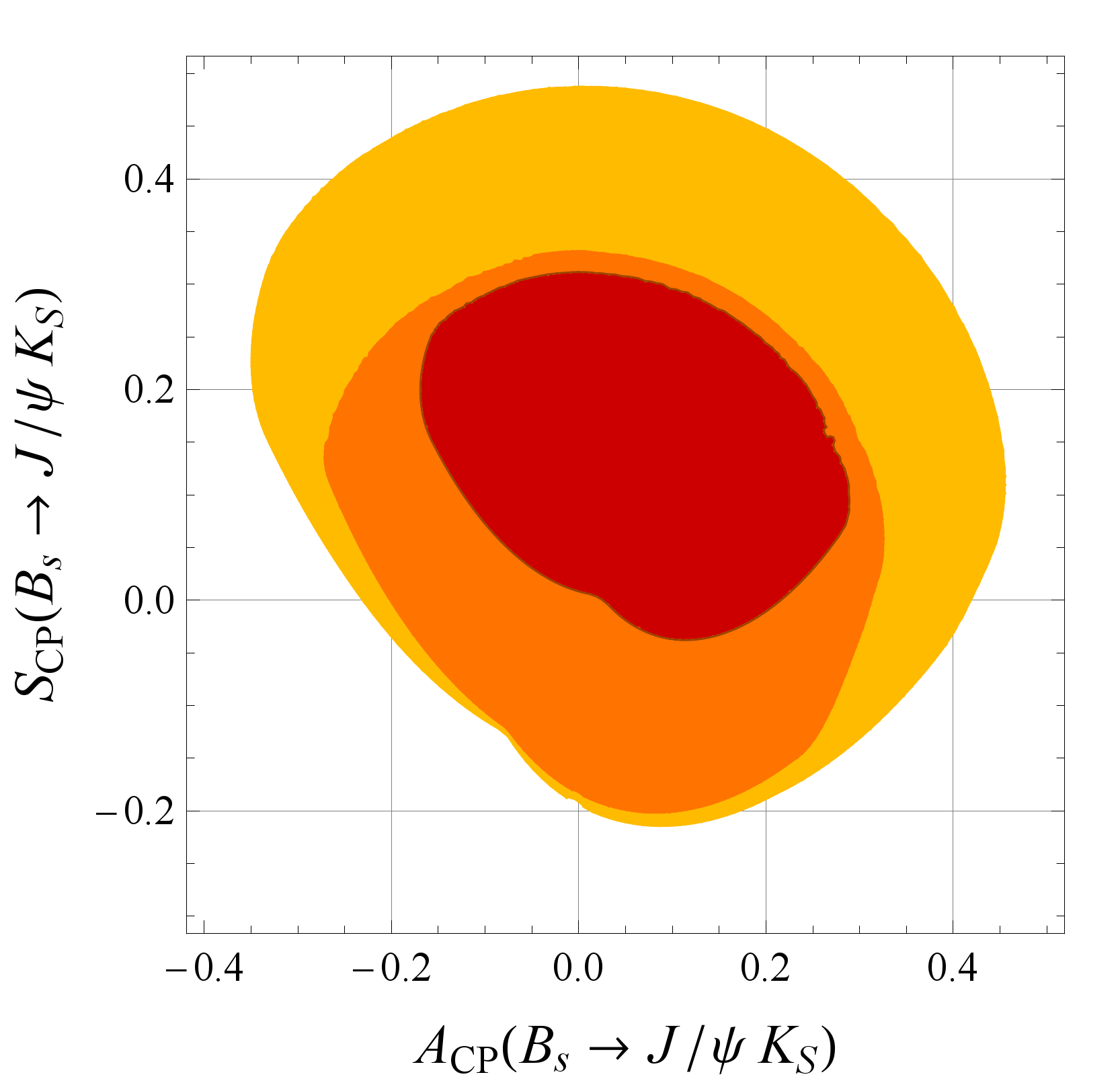}
\caption{\label{fig::resultsfullfitb} Predictions for the two datasets for the CP asymmetries in $B_s\to J/\psi K_S$, including all available data. The colour code is identical to Fig.~\ref{fig::resultsfullfita}.}
\end{center}
\end{figure}

\begin{figure}
\begin{center}
\includegraphics[width=7.8cm]{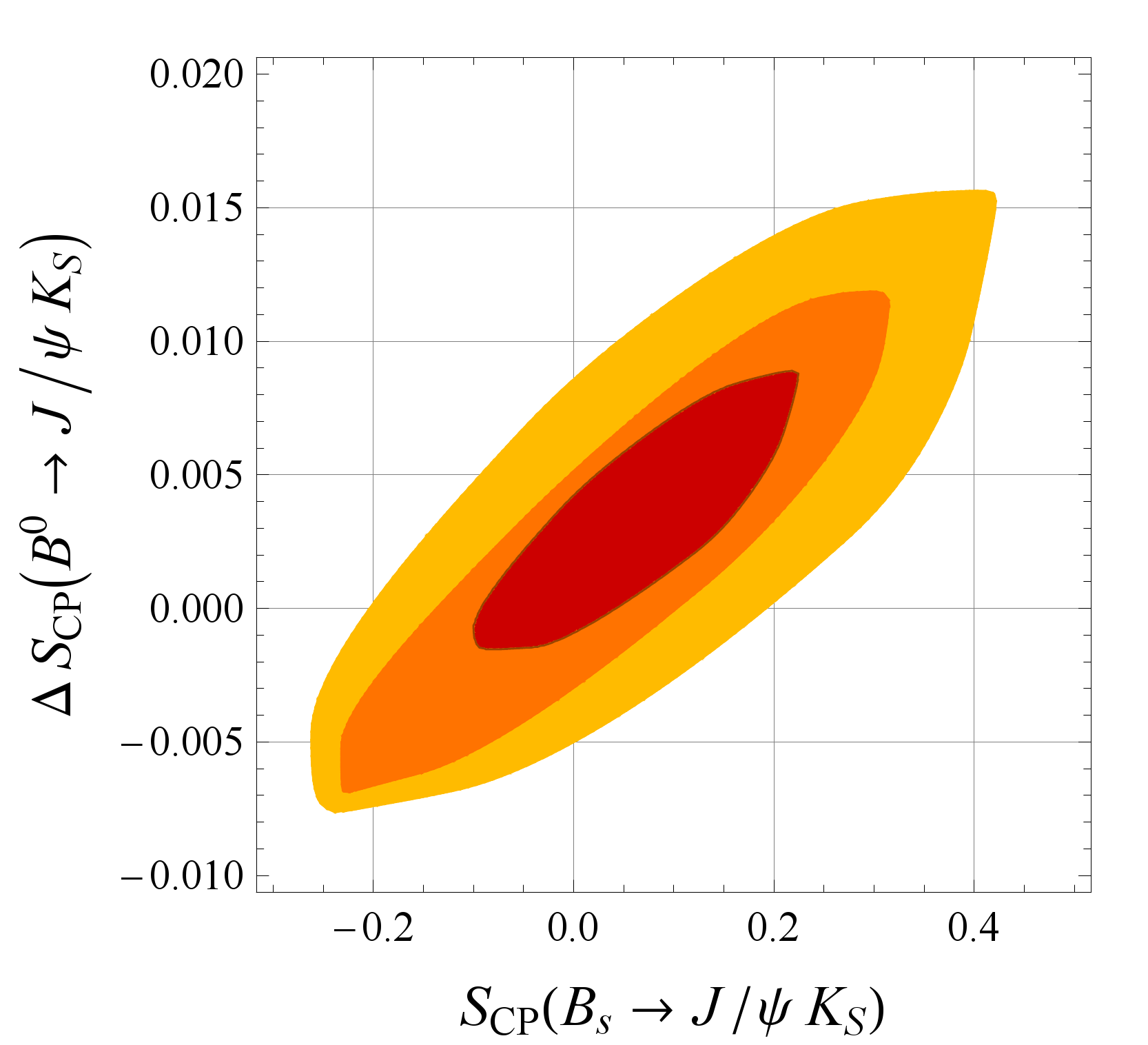}\hfill \includegraphics[width=7.8cm]{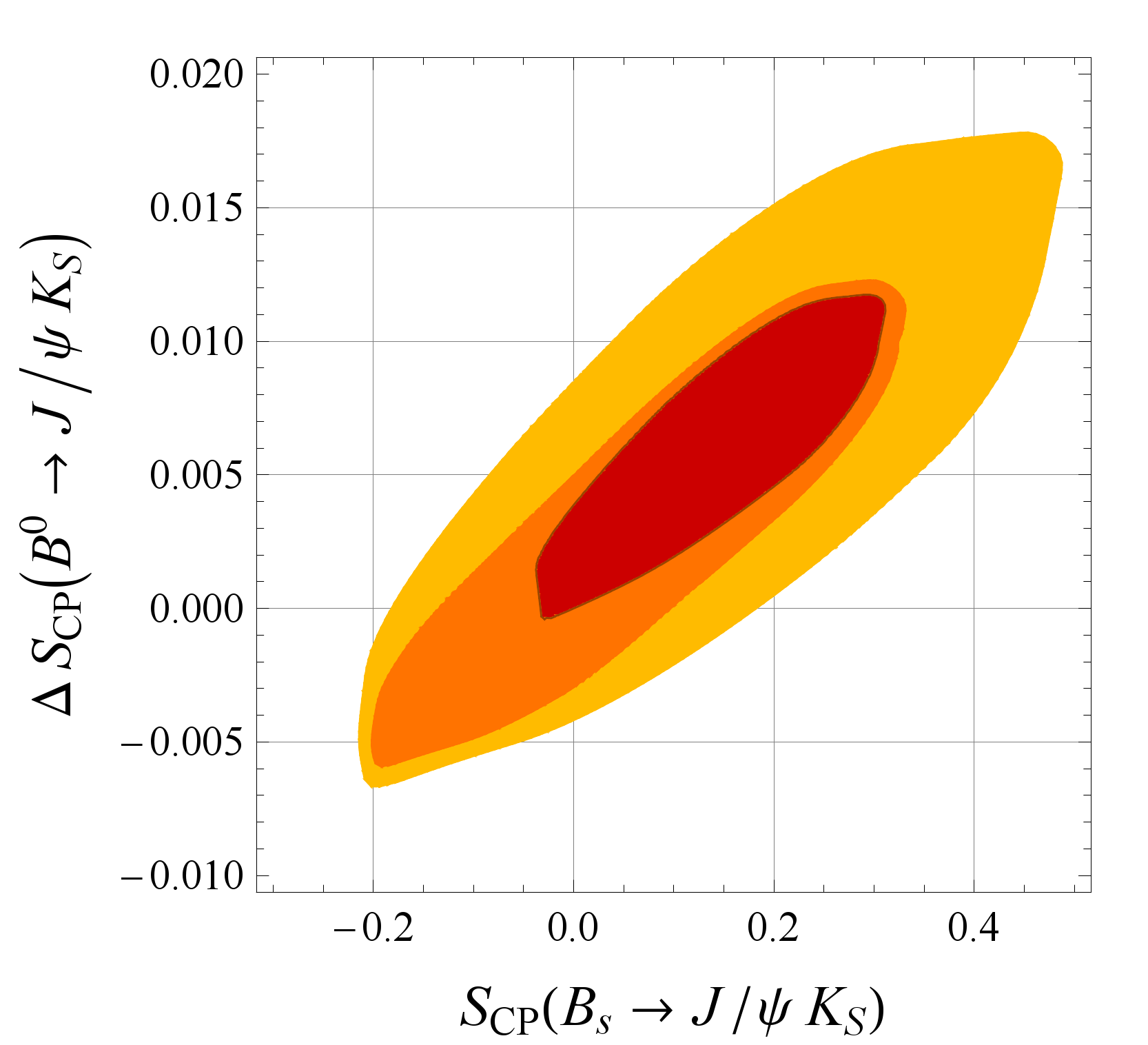}
\caption{\label{fig::resultsfullfitc} Fit results for the two datasets for $\Delta S_{J/\psi K}$ versus $S_{\rm CP}(B_s\to J/\psi K_S)$, including all available data. The colour code is identical to Fig.~\ref{fig::resultsfullfita}.}
\end{center}
\end{figure}

In Fig.~\ref{fig::resultsfullfitb}
the prediction for the CP asymmetries in $B_s\to J/\psi K_S$ is shown for the two datasets. The allowed range remains rather large, which however indicates that the corresponding measurement will add important information to the fit already with moderate precision.
In Fig.~\ref{fig::resultsfullfitc} we plot the correlation between $\Delta S_{J/\psi K_S}$ and $S(B_s\to J/\psi K_S)$, which allows to read off what impact the measurement of only this quantity would have on the phase shift. It will be possible to exclude parts of the allowed range for $\Delta S_{J/\psi K_S}$ with such a measurement; however, to see the full impact of future measurements, an analysis including all expected improvements is needed, which we will show in the next subsection. 

Although in principle the fit is sensitive to the CKM angle $\gamma$ as well (see \cite{FleischerPsiK,DeBruyn:2010hh}), the extracted range from the present data is huge. Actually this is not surprising, given the situation that neglecting all corresponding terms in the amplitudes leads to a rather good fit already. That situation may change, once a  measurement of a significantly non-vanishing direct CP asymmetry is performed.

\subsection{Future scenarios\label{sec::futurescenarios}}
As a main motivation for the present analysis are the expected precision measurements for the decays in question, we analyze their potential impact on the extraction of the weak phases described in the last subsections. For that purpose we define three scenarios, in which we assume improved measurements for some of the observables. Scenario~1 is corresponding to the situation in a few years, with $\sim5~{\rm fb}^{-1}$ of LHCb data.  Another few years later we have scenario~2, with the first $\sim 5~{\rm ab}^{-1}$ of SFF data. Finally scenario 3 corresponds to a situation in the farther future, with $\sim 50~{\rm ab}^{-1}$ of SFF data and $100~{\rm fb}^{-1}$  of Super-LHCb data.
As detailed studies are available only for a few of the observables in question, and a detailed analysis is beyond the scope of this work, we simply scale for the remaining observables the present statistical uncertainties to correspond to the integrated luminosities of the scenario in question. Details to that procedure and the resulting inputs can be found in appendix~\ref{sec::inputsfuture}.

To be able to compare with the results above, and at the same time to estimate the remaining dependence on the theoretical assumptions used in those fits, we add to the plots below another area, showing the allowed range at $95\%$ CL when huge values are allowed for the parameters, i.e. $r_{\rm pen},r_{SU(3)}\lesssim 200\%$. Again, this is for illustration purposes only, to demonstrate the decreasing influence  of the theoretical assumptions with additional data.

\begin{figure}
\begin{center}
\includegraphics[width=7.8cm]{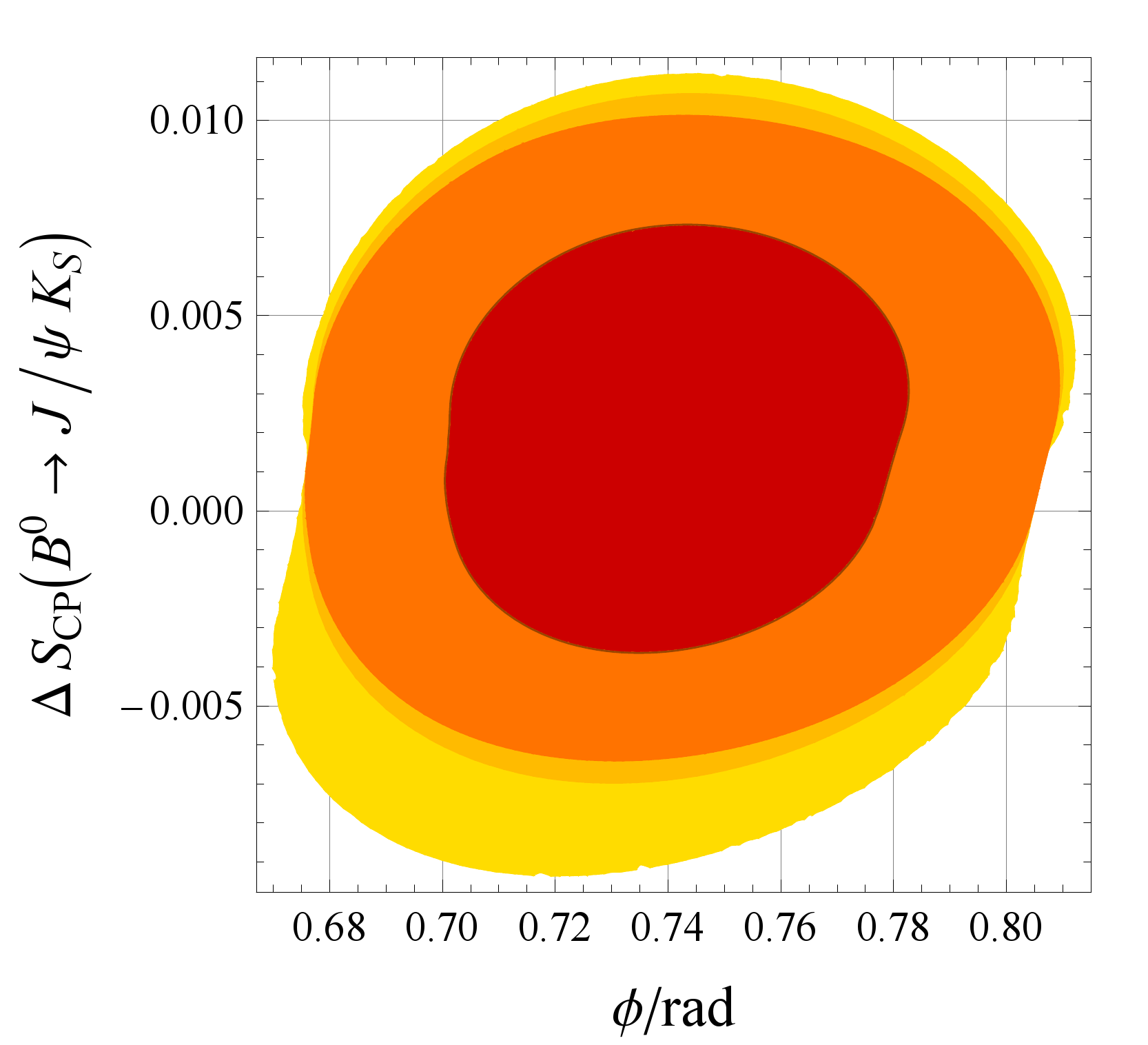}\hfill \includegraphics[width=7.8cm]{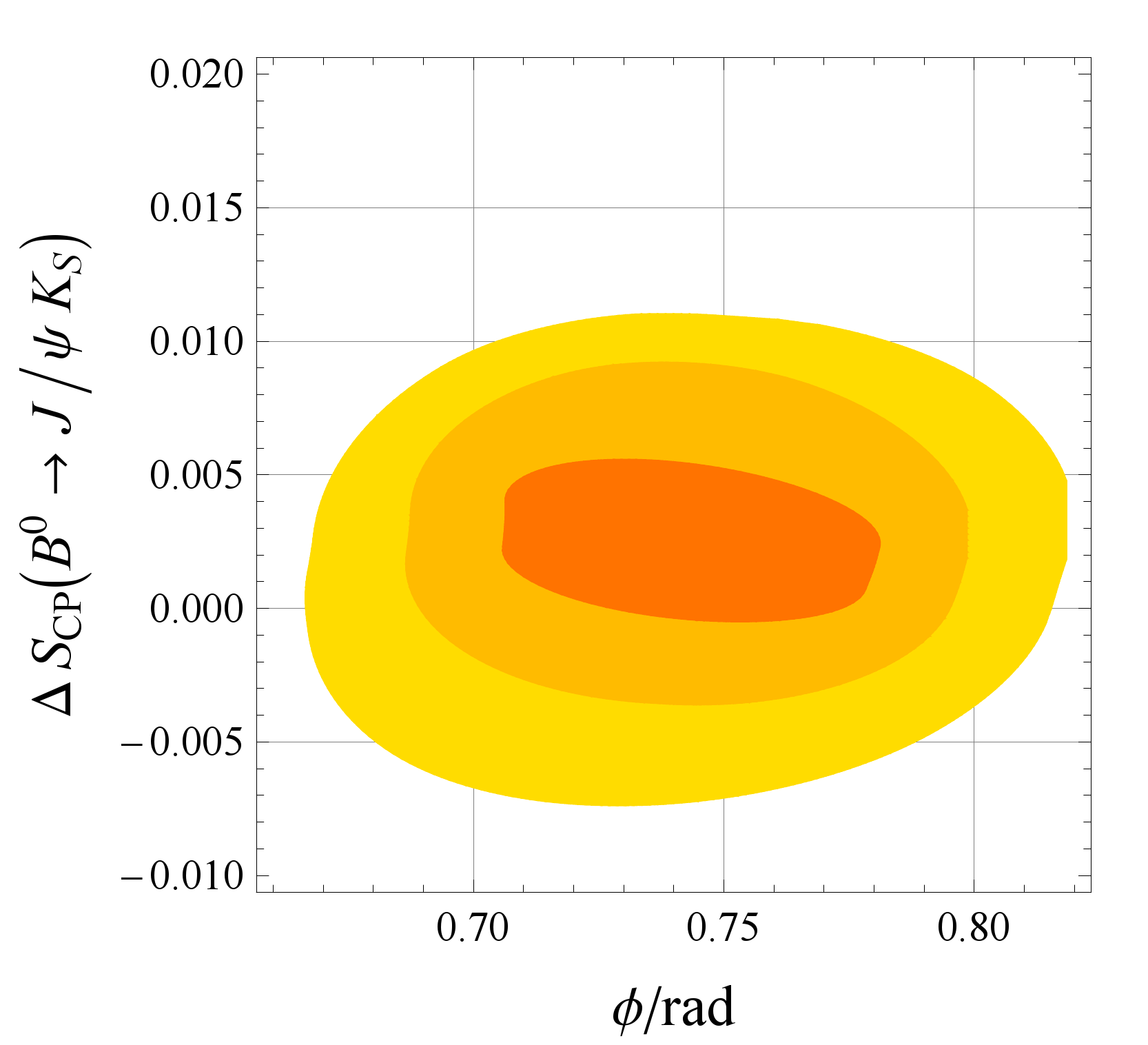}
\caption{\label{fig::resultssc1} Fit results for scenario 1 for $\Delta S_{J/\psi K}$ versus the mixing phase $\phi$ (left), showing the three areas as in  Fig.~\ref{fig::resultsfullfita}, and in addition the $95\%$~CL range with 
$r_{SU(3), {\rm pen}}\sim 200\%$ 
allowed. On the right, we show the $95\%$~CL regions of the three future scenarios, see Sec.~\ref{sec::futurescenarios}, for $r_{SU(3)}=40\%$ and $r_{\rm pen}=50\%$.}
\end{center}
\end{figure}

Starting with scenario~1, the numerical improvement for the uncertainties of the mixing phase $\phi$ and the shift $\Delta S$ is small, as can be seen Tab.~\ref{tab::resultsfuturescs}. However, Fig.~\ref{fig::resultssc1} demonstrates the main effect of the new measurements: when removing the theoretical constraints, the range for $\Delta S$ does not increase much anymore, in contrast to the fit with present data, in which the upper limit grows by almost a factor of three. Although cancellations between some parameters remain possible, the physically relevant combinations are therefore under control with these measurements.
\mytab[tab::resultsfuturescs]{|l|c|c c c|}{\hline
								&	Now		&	Sc.1	&	Sc.2	&	Sc.3	\\\hline
$\delta \phi\,(1\sigma)$		&	$0.030$	&	$0.027$	&	$0.021$	&	$0.015$ \\
$\Delta(\Delta S)\,(95\%)$~CL	&	$0.015$	&	$0.014$	&	
$0.010$	&	$0.005$\\\hline
}
{Results for the errors of the mixing phase $\phi$ and the shift $\Delta S$ ($\Delta(\Delta S)=\Delta S_{\rm max}-\Delta S_{\rm min}$) for the three future scenarios, comparing with the present situation. We do not show the central values, as they correspond to our choice of observables and are arbitrary in that sense.}

Regarding scenarios~2 and 3, the improvement in determining the mixing phase $\phi$ stems from the one in $S(B\to J/\psi K_S)$, and corresponds to the one that would be obtained using the naive relation for the mixing angle with $\Delta S=0$. 
	
Investigating the possible determination of the CKM phase $\gamma$, we observe that the allowed range remains very large ($\pm 35^\circ$), even within scenario 3. The reason for this is twofold: first of all the additional freedom introduced by the $SU(3)$-breaking parameters effectively prevents the penguin contributions to be determined precisely. 
Secondly, even with the precision  for this scenario, some direct CP asymmetries are still compatible with zero, reducing the sensitivity to $\gamma$ further. This of course depends on the central values chosen, but the scenario is close to the present situation. However, to investigate the influence of this choice we repeated the fit with an  
idealized scenario, in which we have small $SU(3)$ breaking and large imaginary parts of the penguin matrix elements. Even in that case we obtain only $\delta\gamma\sim 15^\circ$ for $r_{\rm pen}\leq 40\%$, and that range is only due to the fact that we chose parameters close to this limit, it is much larger without that assumption.

We cannot confirm therefore the optimistic expected ranges for $\gamma$ given in \cite{DeBruyn:2010hh} within this more general framework. Especially the estimate of $SU(3)$-breaking corrections given there does not seem general enough in light of the present analysis; the form factor ratio discussed in that context is \emph{not} the main source of $SU(3)$ breaking in these decays, as emphasized before. Of course, our theoretical understanding of $SU(3)$ might  improve during the next years; however, this kind of development seems hard to quantify.

\subsection{New physics contributions}
In the previous subsections  we established working fits within the SM without a large enhancement for any hadronic parameter, therefore obviously no strong indication for NP in the decay amplitudes is observed. However, as an analysis within a Minimal Flavour Violation framework \cite{D'Ambrosio:2002ex,Chivukula:1987py,Hall:1990ac,Buras:2000dm} yields identical expressions, the possible enhancement of the subleading terms could as well be ascribed to NP contributions. In the latter case, however, additional contributions would be expected in processes with leading contributions from the same terms in the effective hamiltonian, such as $B\to DD$.  

Regarding NP in mixing, the extracted phase 
fits well with the value extracted 
from a UT fit with $|V_{ub}/V_{cb}|$ from semileptonic decays as main input for this quantity, yielding $\phi_{\rm SM}^{\rm sl}=0.84\pm0.13$, as does however the naively extracted value. 

The only actual tension is the one with $B\to\tau\nu$, which is measured well above the expected value \cite{Asner:2010qj,Deschamps:2008de,Bona:2009cj,Lunghi:2009ke,Charles:2011va}. However, it is not clear if this tension should be discussed in the context of NP in mixing (note that the indication of NP in $B_d$ mixing in \cite{Lenz:2012az} is mostly due to this mode, as there the assumption of NP residing \emph{only} in mixing is made): Firstly, this mode has a high sensitivity to NP, especially in form of charged scalars \cite{Grzadkowski:1991kb}. The very recent measurement of the branching ratios of $B\to D^{(*)}\tau\nu$ modes different from the SM predictions \cite{BaBar:2012xj} strengthens this possibility. Secondly, for this tension to be a sign of NP in mixing (only), the extracted values from leptonic and semileptonic decays should coincide, which they do only marginally. Finally, the experimental analysis for 
$B\to\tau\nu$
is extremely difficult. In any case, the observed shift due to penguin contributions is relatively small compared to that difference; it does therefore not offer an explanation for this tension. As mentioned before, it should however be included in the corresponding global analyses of NP in mixing.

\section{Conclusions\label{sec::conclusions}}
The analysis presented here allows to include both, 
penguin and leading $SU(3)$-breaking contributions to the golden mode, in a model-independent way. It thereby improves the extraction of the $B_d$ mixing phase to match the impressive precision for $S(B_d\to J/\psi K_S)$ of present and coming high luminosity colliders. The resulting uncertainty 
equals the one when using the naive relation with $\Delta S_{J/\psi K}=0$, see Fig.~\ref{fig::resultssc1}. 

The fits performed with present data show the importance of $SU(3)$-breaking corrections, without which neither a good fit  nor a reliable value for the phase shift can be obtained. In fact, a reasonable description of the data can at the moment be achieved including \emph{only} these corrections of $\mathcal{O}(20\%)$, with vanishing penguin contributions. However, their inclusion yields an even better description of the data. The resulting phase shift is found to be small, $|\Delta S|\lesssim 0.01$, with a positive sign preferred, thereby slightly reducing the tension between different extractions of the CKM angle $\beta$. This is the most precise determination of these contributions to date.

An important observation is the importance of branching ratio measurements in this ana\-ly\-sis. Especially additional branching ratio measurements in $B\to J/\psi K$ (e.g. the ratio of the charged and neutral modes) would be interesting, given the relatively large central value of the corresponding isospin asymmetry.
Regarding the difference in the $R_{\pi K}$ measurements, the fit prefers the value obtained by LHCb, although this is not conclusive. 

The extraction of the CKM phase $\gamma$ turns out to be more difficult than  expected. One reason lies in the fact that at the moment all direct CP asymmetries are compatible with zero, and, given the present data and projected experimental precision, several of them are likely to remain that way. However, even in an idealized scenario where they are larger, the precision in $\gamma$ remains far from competitive without further knowledge of $SU(3)$ breaking. Neglecting $SU(3)$ breaking (or taking only the one by form factors into account), the pattern of branching ratios allows to determine the subleading terms much more precisely, which is why in that case a more optimistic result was achieved \cite{DeBruyn:2010hh}. 
Of course, a better understanding of these effects might be achieved in the coming years, re-opening this possibility.

The data do not indicate the presence of NP, although the penguin-induced shift in the mixing phase is too small to offer an explanation for the difference to $B\to \tau\nu$. However, the interpretation of this tension in terms of NP in mixing seems questionable, given its sensitivity to NP, and the situation in $B\to D^{(*)}\tau\nu$ and semileptonic decays. A general model-independent analysis of NP in the decay amplitude is impossible; for minimal flavour violating scenarios however the results shown here remain valid, and the enhancement of the penguin contributions compared to existing estimates might be attributed to NP contributions in that case. In this scenario, similar contributions should affect also e.g. $B\to DD$ decays.

\section*{Acknowledgements}
I would like to thank Stefan Schacht for providing part of the code used in the fits and technical support in its adaption. This work
is supported by the Bundesministerium f\"ur Bildung und Forschung (BMBF) under contract No. 05H09PEE.

%===================================================================

\appendix
\renewcommand{\theequation}{\Alph{section}.\arabic{equation}}
\setcounter{equation}{0}

\section{Details of the $\mathbf{SU(3)}$ analysis\label{sec::su3app}}
We use conventions, in which the fundamental triplet is written as
\begin{equation}
(u,d,s)=(\rep{3}_{1/2,1/2,1/3},\rep{3}_{1/2,-1/2,1/3},\rep{3}_{0,0,-2/3})\,.
\end{equation}
For the corresponding antitriplet the $\bar{u}$ has a negative sign, corresponding to the usual isospin representation. The meson states are then defined such that they correspond to ``pure'' $SU(3)$ states without signs, i.e. every meson involving a $\bar{u}$ has an additional sign. This implies the assignments given in Tab.~\ref{tab::finalstates}.

\mytab[tab::finalstates]{|c|c c c c|}{\hline
M           & $\pi^+$   & $\pi^0$   & $\pi^-$    & $\eta_8$\\\hline
$(I,I_z,Y)$ & $(1,1,0)$ & $(1,0,0)$ & $(1,-1,0)$ & $(0,0,0)$\\\hline\hline
M           & $K^+$        & $K^0$          & $\bar{K}^0$    & $K^-$ \\\hline
$(I,I_z,Y)$ &$(1/2,1/2,1)$ & $(1/2,-1/2,1)$ & $(1/2,1/2,-1)$ & $(1/2,-1/2,-1)$\\\hline
}{Assignments of pseudoscalar octet meson states to $SU(3)$ ones.}

The decompositions of the hamiltonian $\mathcal{H}_u$  read, using the isoscalar factors given in \cite{deSwart:1963gc},
\begin{equation}
\mathcal{H}_u^{b\to s}\sim \frac{1}{2}\rep{15}_{1}+\sqrt{\frac{1}{8}}\rep{15}_{0}+\frac{1}{2}\rep{\bar{6}}_{1}+\sqrt{\frac{3}{8}}\rep{3}_{0}
\end{equation} 
with $\Delta I_z=0,\Delta Y=-2/3$ for the $b\to s$ transition, and
\begin{equation}
\mathcal{H}_u^{b\to d}\sim \sqrt{\frac{1}{3}}\rep{15}_{3/2}+\sqrt{\frac{1}{24}}\rep{15}_{1/2}-\frac{1}{2}\rep{\bar{6}}_{1/2}+\sqrt{\frac{3}{8}}\rep{3}_{1/2}
\end{equation}
with $\Delta I_z =-1/2, \Delta Y=1/3$ for the $b\to d$ one. Note that all occuring representations receive contributions from tree- and electroweak-penguin operators, while penguin operators contribute only to triplet matrix elements at this order.

\mytab[tab::su3coeffs]{|l|c c c c|}{\hline
Decay                                   & $\ME{8}{3}_c$    & $\ME{8}{3}_u$         & $\ME{8}{\bar{6}}_u$     & $\ME{8}{15}_u$\\\hline
$\bar{B}_d\to J/\psi \bar{K}^0$         & $1$              & $\sqrt{\frac{3}{8}}$  & $\sqrt{\frac{1}{12}}$   & $\sqrt{\frac{1}{120}}$\\
$\bar{B}_d\to J/\psi \pi^0$             & $1/\sqrt{2}$     & $\sqrt{\frac{3}{16}}$ & $-\sqrt{\frac{1}{24}}$  & $\sqrt{\frac{5}{48}}$\\
$B^-\to J/\psi K^-$                     & $1$              & $\sqrt{\frac{3}{8}}$  & $-\sqrt{\frac{1}{12}}$  & $-\sqrt{\frac{3}{40}}$\\
$B^-\to J/\psi \pi^-$                   & $1$              & $\sqrt{\frac{3}{8}}$  & $-\sqrt{\frac{1}{12}}$  & $-\sqrt{\frac{3}{40}}$\\
$\bar{B}_s\to J/\psi \pi^0$             & $0$              & $0$                   & $-\sqrt{\frac{1}{6}}$   & $\sqrt{\frac{1}{15}}$\\
$\bar{B}_s\to J/\psi K^0$               & $1$              & $\sqrt{\frac{3}{8}}$  & $\sqrt{\frac{1}{12}}$   & $\sqrt{\frac{1}{120}}$\\
\hline
}{Coefficients for $B\to J/\psi P$ amplitudes in the $SU(3)$ limit.}

Projecting these operators on the relevant initial and final states yields the coefficients of the corresponding  reduced matrix elements 
given in Tab.~\ref{tab::su3coeffs}. Note that the amplitudes obey the trivial $U$-spin relations for $B^-\to J/\psi P^-$ (P=$\pi,K$) and between $\bar{B}^0\to J/\psi \bar{K}^0$ and $\bar{B}_s\to J/\psi K^0$, as well as isospin relations between the neutral and charged modes in $B\to J/\psi K$ and $B\to J/\psi\pi$. We confirm the expressions obtained in \cite{Zeppenfeld}\footnote{In \cite{Zeppenfeld}, a different normalization for the reduced matrix elements in $A_u$ is used. The translation reads $\ME{8}{15}_u^Z=-\sqrt{1/24}\ME{8}{15}_u$, $\ME{8}{\bar{6}}_u^Z=\ME{8}{\bar{6}}/2$, and $\ME{8}{3}_u^Z=\sqrt{3/8}\ME{8}{3}_u$.}.

Regarding $SU(3)$ breaking, the tensor product of the leading-order hamiltonian $\mathcal{H}_c$ with the isospin-conserving breaking term leads to
\begin{eqnarray}
\mathcal{H}_c^{b\to s}\otimes \rep{8}_{0,0,0}&\sim& \sqrt{\frac{3}{4}}\rep{15}_{0,0,-2/3}^\epsilon+\frac{1}{2}\rep{3}_{0,0,-2/3}^\epsilon\,,\quad{\rm and}\nonumber\\
\mathcal{H}_c^{b\to d}\otimes \rep{8}_{0,0,0}&\sim& \frac{3}{4}\rep{15}_{1/2,-1/2,1/3}^\epsilon-\sqrt{\frac{3}{8}}\rep{\bar{6}}_{1/2,-1/2,1/3}-\frac{1}{4}\rep{3}_{1/2,-1/2,1/3}^\epsilon\,,
\end{eqnarray}
implying three additional reduced matrix elements, whose coefficients are given in Tab.~\ref{tab::su3coeffseps}. 

\mytab[tab::su3coeffseps]{|l|c c c|}{\hline
Decay                           & $\ME{8}{3}_c^\epsilon$ & $\ME{8}{\bar{6}}_c^\epsilon$  & $\ME{8}{15}_c^\epsilon$\\\hline
$\bar{B}_d\to J/\psi \bar{K}^0$ & $\frac{1}{2}$          & $0$                           & $-\sqrt{\frac{1}{20}}$\\
$\bar{B}_d\to J/\psi \pi^0$     & $-\sqrt{\frac{1}{32}}$ & $-\frac{1}{4}$                & $-\sqrt{\frac{1}{160}}$\\
$B^-\to J/\psi K^-$             & $\frac{1}{2}$          & $0$                           & $-\sqrt{\frac{1}{20}}$\\
$B^-\to J/\psi \pi^-$           & $-\frac{1}{4}$         & $-\sqrt{\frac{1}{8}}$         & $-\sqrt{\frac{1}{80}}$\\
$\bar{B}_s\to J/\psi \pi^0$     & $0$                    & $0$                           & $0$\\
$\bar{B}_s\to J/\psi K^0$       & $-\frac{1}{4}$         & $\sqrt{\frac{1}{8}}$          & $\sqrt{\frac{9}{80}}$\\
\hline
}{Coefficients for the linear $SU(3)$-breaking amplitudes from $\mathcal{H}_c$ in $B\to J/\psi P$ decays.}

The corresponding coefficient matrix does not have maximal rank, i.e. it is possible to express the corresponding part of the amplitudes with less parameters, for example choosing the combinations $A_{\epsilon 1}=\ME{8}{3}_c^\epsilon-\sqrt{1/5}\ME{8}{15}_c^\epsilon$ and $A_{\epsilon 2}=\ME{8}{\bar{6}}_c^\epsilon+\sqrt{2/5}\ME{8}{15}_c^\epsilon$. Note that despite the somewhat small coefficient of the second matrix element in these relations, it is of course not clear which one will be dominating these combinations (if any). The dominance of one of the first ones would be signalled by only the corresponding one being sizable, while dominance of the absorbed one would result in $-\sqrt{2}A_{\epsilon 1}\simeq A_{\epsilon 2}$ in the following fits.

For the parametrization given in Eq.~(\ref{eq::parametrization}) we introduced the following abbreviations:\newline $\mathcal{N}=V_{cb}V_{cs}^*\ME{8}{3}_c$ is the leading amplitude which is chosen to be real. The other matrix elements are normalized to the leading one,  
\begin{equation}
R_{\epsilon 1}=A_{\epsilon1}/(4\ME{8}{3}_c) \mbox{\quad and \quad}R_{\epsilon 2}=A_{\epsilon 2}/(2\sqrt{2}\ME{8}{3}_c)
\end{equation}
including the $SU(3)$-violating combinations  
of matrix elements introduced before, 
\begin{equation}
R_{u1} = R_u\frac{\sqrt{45}\ME{8}{3}_u-\ME{8}{15}_u}{\sqrt{120}\ME{8}{3}_c} \mbox{\quad and \quad}R_{u2}=R_u\frac{\sqrt{2}\ME{8}{15}_u}{\sqrt{15}\ME{8}{3}_c}
\end{equation} 
the dominant subleading contributions, and  the remaining one  
\begin{equation}
\delta=R_u\frac{\sqrt{5}\ME{8}{6}_u-\sqrt{2}\ME{8}{15}_u}{\sqrt{60}\ME{8}{3}_c}\,,
\end{equation} 
which is expected to be even smaller.
Allowing for $r_{SU(3)}\equiv|A_{\epsilon1,2}|/|\ME{	8}{3}_c|\leq40\%$ implies $|R_{\epsilon1}|\leq 10\%$ and $|R_{\epsilon2}|\leq14\%$. Furthermore we impose in the constrained fit scenarios $r_{\rm pen}=|\ME{8}{N}_u|/|\ME{8}{3}_c|\leq50\%$, 
to be compared with $r_{\rm pen}\leq8\%$ from \cite{Boos,Li,GronauRosnersuppressedterms}.
Note that this parametrization corresponds to the one given in \cite{Jung:2009pb} for the subclass of $B_{u,d}$ decays, although the structure of the breaking terms looks different. The reason is that the $SU(3)$-breaking term considered here, although conceptionally identical to one considered in that paper, includes a $\Delta U=0$ part, which can be absorbed for $U$-spin-related decays in the leading amplitude. Furthermore, the limit $\delta\to0$ corresponds to $r_1=r_{3/2}$ and $\phi_1=\phi_{3/2}$ in that paper.

\section{Inputs for future scenarios\label{sec::inputsfuture}}
We collect the estimated inputs for the different future scenarios discussed in Sec.~\ref{sec::futurescenarios}. To find them, we proceed as follows: Explicit analyses are used where available; when they are absent, but a measurement is available from the corresponding experiment, the statistical uncertainties are scaled to correspond to the luminosity in the scenario considered, while leaving the systematic errors untouched, i.e. $\sigma_{\rm now}=\sqrt{\sigma_{\rm stat}^2+\sigma_{\rm syst}^2}\to \sigma_{\rm future}=\sqrt{\sigma_{\rm stat}^2L_{\rm now}/L_{\rm future}+\sigma_{\rm syst}^2}$. We proceed the same way  for $B$-factory results to extract the expected SFF measurements, although this results in an even cruder estimate. Finally, when no measurement exists yet, we ``guesstimate'' the order of magnitude for the corresponding uncertainties, trying to stay on the conservative side. The listed uncertainties correspond to only one experiment, although of course in many cases different experiments will be able to measure them, thereby reducing the uncertainties further. Exceptions are the branching ratios in $B\to J/\psi K$, for which we assume some progress despite them being already dominated by systematic errors, and the corrsponding CP asymmetries in scenario 1, where the number given corresponds to a guessed new world average. The results of this procedure are shown in Tab.~\ref{tab::expdatafuture}. 
\begin{table}[thb]
\begin{center}
\resizebox{\textwidth}{!}{
\begin{tabular}{|l|l|c|c c c|c|}\hline
Observable                      & ${\mathcal{O}}_{\rm future}$ & $\delta\mathcal{O}_{\rm now}$     & $\delta \mathcal{O}_{\rm Sc1}$ & $\delta \mathcal{O}_{\rm Sc2}$ & $\delta \mathcal{O}_{\rm Sc3}$ & Ref.\\\hline
$BR(\bar{B}^0\to J/\psi \bar{K}^0)/10^{-4}$	& $\phantom{-}8.90$		& $0.32$    & $0.25$	& $0.20$	& $0.20$	& our guess (WA)\\
$A_{\rm CP}(\bar{B}^0\to J/\psi \bar{K}^0)$	& $\phantom{-}0.012$	& $0.021$	& $0.019$ 	& $0.015$	& $0.013$	& our guess (WA) / \cite{Aushev:2010bq}\\
$S_{\rm CP}(B\to J/\psi K_S)$				& $\phantom{-}0.672$	& $0.022$	& $0.020$ 	& $0.016$	& $0.012$	& our guess (WA) / \cite{Aushev:2010bq}\\
$BR(\bar{B}^0\to J/\psi \pi^0)/10^{-4}$		& $\phantom{-}0.178$	& $0.016$	& $0.016$	& $0.008$	& $0.007$	& our estimate\\
$A_{\rm CP}(\bar{B}^0\to J/\psi \pi^0)$		& $\phantom{-}0.077$	& $0.13$	& $0.13$	& $0.06$	& $0.021$	& chosen equal to $S_{\rm CP}$\\
$S_{\rm CP}(B\to J/\psi \pi^0)$				& $-0.740$				& $0.15$	& $0.15$	& $0.06$	& $0.021$	& \cite{O'Leary:2010af}\\
$BR(B^-\to J/\psi K^-)/10^{-4}$	    		& $\phantom{-}9.68$		& $0.34$	& $0.25$	& $0.20$	& $0.20$	& our guess (WA)\\
$A_{\rm CP}(B^-\to J/\psi K^-)$				& $\phantom{-}0.000$	& $0.007$	& $0.007$ 	& $0.003$	& $0.002$	& our estimate\\
$R_{\pi K}/\%$								& $\phantom{-}3.78$		& $0.13$	& $0.08$	& $0.08$	& $0.07$	& our estimate \\%Estimate from SuperB weaker
$A_{\rm CP}(B^-\to J/\psi \pi^-)$			& $\phantom{-}0.004$	& $0.029$	& $0.013$	& $0.012$	& $0.005$	& our estimate\\	 %$0.011$ Sc3 with LHCb\\
$R_{KK}/\%$ 						      	& $\phantom{-}3.67$		& $0.4$		& $0.25$	& $0.25$	& $0.22$	& our estimate\\
$A_{\rm CP}(\bar{B}_s\to J/\psi K^0)$  		& $\phantom{-}0.09$		& ---		& $0.1$ 	& $0.1$		& $0.05$	& our guess\\
$S_{\rm CP}(B_s \to J/\psi K_S)$			& $\phantom{-}0.01$		& ---		& $0.1$		& $0.1$		& $0.05$	& our guess\\\hline
\end{tabular}
}
\end{center}
\caption{\label{tab::expdatafuture}Experimental data used for the analysis of the future scenarios 1-3. ${\mathcal{O}}_{\rm future}$ denotes the central value assumed for all scenarios.} 
\end{table}
In order to create acceptable fits, we choose the projected central values to lie gaussian distributed around the values of an idealized scenario similar to dataset 2, with the uncertainties from future scenario 3.

\bibliography{BtoJPsiP}
\end{document}